\documentclass[10pt,conference]{IEEEtran}

\def\istechreport{}

\usepackage{cite}
\usepackage{amsmath,amssymb,amsfonts,amsthm}
\usepackage{algorithmic}
\usepackage{graphicx}
\usepackage{textcomp}
\usepackage{xcolor}

\usepackage{helpers}
\usepackage{caches}
\usepackage{mytikz} 
\usepackage{myhyphen}
\usepackage{notation}

\usepackage{thm-restate}

\usepackage{hyperref}

\usepackage{bm}

\usepackage{microtype}
\usepackage{paralist}

\newcommand\spacedstyle[1]{\SetTracking{encoding=*}{#1}\lsstyle}

\usepackage{balance}

\usepackage{booktabs}

\usepackage{xspace}
\usepackage{xparse}

\newcommand\etal{et~al.\xspace}

\usepackage{verbatim}

\NewDocumentEnvironment{techreport}{}
{ %
    \ifthenelse{\isundefined{\istechreport}}%
    {%
       \comment%
    }{%
    }%
}{ %
    \ifthenelse{\isundefined{\istechreport}}%
    {%
    }{%
    }%
}

\NewDocumentEnvironment{paperonly}{}
{%
    \ifthenelse{\isundefined{\istechreport}}%
    {%
    }{%
        \comment%
    }%
}{%
    \ifthenelse{\isundefined{\istechreport}}%
    {%
    }{%
    }%
}

\newcommand{\refthms}[1]{(\text{Th.}~\ref{thm:#1})}
\newcommand{\refcors}[1]{(\text{Cor.}~\ref{cor:#1})}

\newcommand{\llvmta}{\textsc{llvmta}\xspace}

\renewcommand{\todo}[1]{}

\begin{document}
\newcommand{\graphwcetmru}{\begin{figure*}[ht!]
\centering
	\input{rtas_graphs/WCET_for_selected_programs.pgf}
	\vspace{-5mm}
	\caption{WCET bounds for different competitiveness-based analyses for \mru ($k = 4$, $\textit{size} = 4$~KB) normalized to the \emph{\lru may/must+persistence analysis}.\label{fig:ex_block_vs_block_miss_nmru}}
	\vspace{-1mm}
\end{figure*}
}
\newcommand{\graphwcetfifo}{\begin{figure*}[ht!]
	\centering
	\input{rtas_graphs/WCET_for_selected_programs_fifo.pgf}
	\vspace{-5mm}
	\caption{WCET bounds for different competitiveness-based analyses for \fifo ($k = 4$, $\textit{size} = 4$~KB) normalized to the \emph{\lru may/must+persistence analysis}.\label{fig:ex_block_vs_block_miss_fifo}}
	\vspace{-1mm}
\end{figure*}
}
\newcommand{\graphmisses}{\begin{figure}[ht!]
	\centering
	\input{rtas_graphs/Misses_for_different_assotiativities.pgf}
	\vspace{-5mm}
	\caption{Maximal number of misses for the best analyses for \lru, \mru, and \fifo for different associativities and 64 cache sets. The results are normalized to the \lru \emph{must/may+persistence analysis} for associativity $k = 1$ and summarized across all benchmarks via the geometric mean.\label{fig:acc_over_k}}
	\vspace{-1mm}
\end{figure}}
\newcommand{\graphwcetsize}{\begin{figure}[ht!]
	\centering
	\input{rtas_graphs/WCET_with_varying_size_k=4.pgf}
	\vspace{-5mm}
	\caption{WCET bounds for the best analyses for \lru, \mru, and \fifo for different cache sizes at a fixed associativity of $k=4$ normalized to \lru \emph{must/may+persistence analysis} for size $0.5$~KB and summarized across all benchmarks via the geometric mean.\label{fig:acc_over_s}}
	\vspace{-1mm}
\end{figure}}
\newcommand{\graphruntime}{\begin{figure}[ht!]
	\centering
	\input{rtas_graphs/Runtime_with_varying_size_k=4.pgf}
	\vspace{-6mm}
	\caption{Cost of different analysis for different cache sizes (all with $k = 4$) normalized to \lru \emph{must/may+persistence analysis} for size $0.5$~KB and summarized across all benchmarks via the geometric mean.\label{fig:runtime_over_s}}
	\vspace{-1mm}
\end{figure}}
\newcommand{\graphcosts}{\begin{figure}[ht!]
	\centering
	\input{rtas_graphs/cost_for_different_analyses.pgf}
	\vspace{-6mm}
	\caption{Cost of different analyses for a cache with $k = 4$ and $\textit{size} = 4$~KB normalized to the LRU may/must + persistence analysis and summarized across all benchmarks via the geometric mean.\looseness=-1\label{fig:cost_diff_analysis}}
	\vspace{-1mm}
\end{figure}}

\title{A Unified Framework for\\Quantitative Cache Analysis}

\author{\IEEEauthorblockN{Sophie Kahlen and Jan Reineke}
\IEEEauthorblockA{
\textit{Saarland University}\\
\textit{Saarland Informatics Campus} \\
Saarbrücken, Germany
}
}

\maketitle
\thispagestyle{plain}
\pagestyle{plain}

\begin{abstract}
	In this work we unify two existing lines of work towards cache analysis for non-LRU policies.
    To this end, we extend the notion of competitiveness to block competitiveness and systematically analyze the competitiveness and block competitiveness of FIFO and MRU relative to LRU for arbitrary associativities.
	We show how competitiveness and block competitiveness can be exploited in state-of-the-art WCET analysis based on the results of existing persistence analyses for LRU.
	Unlike prior work, our approach is applicable to microarchitectures that exhibit timing anomalies.
	We experimentally evaluate the precision and cost of our approach on benchmarks from TACLeBench.
	The experiments demonstrate that quantitative cache analysis for FIFO and MRU comes close to the precision of LRU.
\end{abstract}

\begin{IEEEkeywords}
cache analysis, persistence analysis, WCET analysis, FIFO, NMRU, implicit path enumeration
\end{IEEEkeywords}

\section{Introduction}

Real-time systems need to satisfy timing constraints imposed by the physical environment in which the system operates.
Determining a program's worst-case execution time (WCET) is a key step in proving that a real-time system meets its timing constraints.\looseness=-1

WCET analysis is challenging due to the timing variability introduced by performance-enhancing features of modern microarchitectures, such as caches, pipelines, and branch predictors.
In this paper, we focus on the analysis of a program's cache behavior.

Every instruction fetch and every data load results in either a cache hit or a cache miss, where the latter introduces a significant performance penalty as the data has to be fetched from a slower memory level.
Thus, accurate cache analysis is crucial to obtain accurate bounds on a program's WCET.

Two kinds of cache analyses are used in WCET analysis: 
\begin{itemize}
	\item \emph{Classifying analyses} classify memory accesses as \emph{always hit}, \emph{always miss}, or as \emph{unknown}.
	\item \emph{Persistence analyses} determine whether blocks are \emph{persistent} in the cache once they have been accessed, which implies that they may cause at most one cache miss.
\end{itemize}

Accurate classifying~\cite{Alt1996,Chattopadhyay2013,Touzeau2017,Touzeau2019,Brandner2022} and persistence analyses~\cite{Arnold94,Mueller94,White97,Ferdinand97,Ferdinand1999rts,Mueller2000rts,Ballabriga2008,Cullmann2011,Huynh2011,Nagar2012,Nagar2012thesis,Cullmann2013tecs,Cullmann2013thesis,Zhang15,Stock2019,Reineke2018b} have been developed for the \emph{least-recently-used} (\lru) replacement policy.
It has been argued in the literature that \lru is the most predictable cache replacement policy~\cite{Reineke07} and that classifying analyses are bound to be less accurate for other policies.\looseness=-1

In practice, however, other policies such as \emph{first-in, first-out} (\fifo), \emph{not most-recently-used} (\mru)~\cite{Eklov11}, \emph{pseudo-LRU}~(\plru)~\cite{al04}, or quad-age LRU (\qlru)~\cite{jaleel10,jahagirdar12} are often preferred over \lru as they exhibit other advantages, such as better average-case performance or lower energy consumption.
Due to the prevalence of these policies in real-world systems, it is important to develop accurate analyses for them.\looseness=-1

There are two promising lines of prior work tackling non-\lru policies in the literature:
\begin{enumerate}
	\item Reineke and Grund~\cite{Reineke08} introduced the notion of \emph{hit} and \emph{miss competitiveness} of pairs of policies. %
	      Based on this notion, they derived classifying analyses for \fifo, \mru, and \plru from classifying analyses for \lru.	      
	\item Guan, Lv, Yang, Yu, and Yi~\cite{Guan12,Guan13,Guan14} determined conditions that allow to lower bound the number of hits for \fifo and to upper bound the number of misses for \mru.
		  They also showed how to exploit these conditions based on the results of existing persistence analyses for \lru.\looseness=-1
\end{enumerate}

\newcommand{\miss}{$\bullet$}
\newcommand{\hit}{$\circ$}

\begin{figure}
	\centering
	\begin{tikzpicture}
		\node[draw, gray, inner sep=1mm] at (0,0) { 
			\begin{tikzpicture}[node distance=0.576cm, black]%
		\node (sigma) {}; %
		
		\node[right of=sigma, xshift=5mm] (a1) {\strut $a$};
		\node[right of=a1] (a2) {\strut $\bm{b}$};
		\node[right of=a2] (a3) {\strut $c$};
		\node[right of=a3] (a4) {\strut $c$};
		\node[right of=a4] (a5) {\strut $\bm{b}$};
		\node[right of=a5] (a6) {\strut $d$};
		\node[right of=a6] (a7) {\strut $\bm{b}$};
		\node[right of=a7] (a8) {\strut $e$};
		\node[right of=a8] (a9) {\strut $\bm{b}$};
		\node[right of=a9] (a10) {\strut $f$};
		\node[right of=a10] (a11) {\strut $f$};
		\node[right of=a11] (a12) {\strut $\bm{b}$};
		
		\node[below of=sigma, yshift=1mm] (lru) {\strut $\lru(2)$:};
		\node[below of=a1, yshift=1mm] (h1) {\strut \miss};
		\node[below of=a2, yshift=1mm] (h2) {\strut \miss};
		\node[below of=a3, yshift=1mm] (h3) {\strut \miss};
		\node[below of=a4, yshift=1mm] (h4) {\strut \hit}; 
		\node[below of=a5, yshift=1mm] (h5) {\strut \hit};
		\node[below of=a6, yshift=1mm] (h6) {\strut \miss}; 
		\node[below of=a7, yshift=1mm] (h7) {\strut \hit};
		\node[below of=a8, yshift=1mm] (h8) {\strut \miss};
		\node[below of=a9, yshift=1mm] (h9) {\strut \hit};
		\node[below of=a10, yshift=1mm] (h10) {\strut \miss}; 
		\node[below of=a11, yshift=1mm] (h11) {\strut \hit};   
		\node[below of=a12, yshift=1mm] (h12) {\strut \hit};   

		\node[below of=lru, yshift=1mm] (fifo) {\strut $\fifo(2)$:};
		\node[below of=h1, yshift=1mm] (f1) {\strut \miss};
		\node[below of=h2, yshift=1mm] (f2) {\strut \miss};
		\node[below of=h3, yshift=1mm] (f3) {\strut \miss};
		\node[below of=h4, yshift=1mm] (f4) {\strut \hit}; 
		\node[below of=h5, yshift=1mm] (f5) {\strut \hit};
		\node[below of=h6, yshift=1mm] (f6) {\strut \miss}; 
		\node[below of=h7, yshift=1mm] (f7) {\strut \miss};
		\node[below of=h8, yshift=1mm] (f8) {\strut \miss};
		\node[below of=h9, yshift=1mm] (f9) {\strut \hit};
		\node[below of=h10, yshift=1mm] (f10) {\strut \miss}; 
		\node[below of=h11, yshift=1mm] (f11) {\strut \hit};   
		\node[below of=h12, yshift=1mm] (f12) {\strut \miss};   

		\node[below of=fifo, yshift=1mm] (blocklru) {\strut $\lru(2),\bm{b}$:};
		\node[below of=f1, yshift=1mm] (bl1) {\strut };
		\node[below of=f2, yshift=1mm] (bl2) {\strut \miss};
		\node[below of=f3, yshift=1mm] (bl3) {\strut };
		\node[below of=f4, yshift=1mm] (bl4) {\strut }; 
		\node[below of=f5, yshift=1mm] (bl5) {\strut \hit};
		\node[below of=f6, yshift=1mm] (bl6) {\strut }; 
		\node[below of=f7, yshift=1mm] (bl7) {\strut \hit};
		\node[below of=f8, yshift=1mm] (bl8) {\strut };
		\node[below of=f9, yshift=1mm] (bl9) {\strut \hit};
		\node[below of=f10, yshift=1mm] (bl10) {\strut }; 
		\node[below of=f11, yshift=1mm] (bl11) {\strut };   
		\node[below of=f12, yshift=1mm] (bl12) {\strut \hit}; 

		\node[below of=blocklru, yshift=1mm] (blockfifo) {\strut $\fifo(2),\bm{b}$:};
		\node[below of=bl1, yshift=1mm] (bf1) {\strut };
		\node[below of=bl2, yshift=1mm] (bf2) {\strut \miss};
		\node[below of=bl3, yshift=1mm] (bf3) {\strut };
		\node[below of=bl4, yshift=1mm] (bf4) {\strut }; 
		\node[below of=bl5, yshift=1mm] (bf5) {\strut \hit};
		\node[below of=bl6, yshift=1mm] (bf6) {\strut }; 
		\node[below of=bl7, yshift=1mm] (bf7) {\strut \miss};
		\node[below of=bl8, yshift=1mm] (bf8) {\strut };
		\node[below of=bl9, yshift=1mm] (bf9) {\strut \hit};
		\node[below of=bl10, yshift=1mm] (bf10) {\strut }; 
		\node[below of=bl11, yshift=1mm] (bf11) {\strut };   
		\node[below of=bl12, yshift=1mm] (bf12) {\strut \miss}; 
	\end{tikzpicture}};
	\end{tikzpicture}
	\caption{Sequence of memory accesses and resulting hits~(\hit) and misses~(\miss) for $\lru(2)$ and $\fifo(2)$, followed by hits and misses focused on block $b$.\label{fig:example}}
\end{figure}

In this paper, we unify and generalize these two lines of work. %
To this end, we introduce the notion of \emph{block-hit} and \emph{block-miss competitiveness}, which relate the performance of different replacement policies focussing on the hits and misses incurred by accesses to an arbitrary designated memory block.
Block competitiveness allows to express the key conditions identified by Guan~\etal~\cite{Guan12,Guan13,Guan14} in a uniform way and can be seen as an extension of \emph{hit} and \emph{miss competitiveness}~\cite{Reineke08}, which relate the total numbers of hits and misses of two policies.

As an example, consider \reffig{example}, which shows a sequence of memory accesses and the resulting hits and misses from initially empty caches of capacity $2$ under $\lru$ and $\fifo$ replacement.
There are five accesses to block $b$.
In the $\lru$ cache, only the first access to $b$ misses, while all subsequent accesses hit, \ie, block~$b$ is \emph{persistent} in the cache.
On the other hand, in the $\fifo$ cache every other access to $b$ misses.
Indeed we show that $\fifo$ is \emph{not} block-miss competitive relative to $\lru$, meaning that a block that is persistent in $\lru$ can experience an arbitrary number of misses in $\fifo$.
However, we also show that $\fifo(2)$ is $\frac{1}{2}$-block-hit competitive relative to $\lru(2)$, meaning that accesses to any block experience at least half the number of hits in $\fifo(2)$ compared with $\lru(2)$.\looseness=-1

As our primary contribution, we systematically analyze the hit and miss competitiveness as well as the block-hit and block-miss competitiveness of \fifo and \mru relative to \lru for arbitrary associativities.

Our second contribution is to show how both competitiveness and block competitiveness can be exploited within \emph{implicit path enumeration}~\cite{Li1995} in state-of-the-art WCET analysis.
As Guan et al.~\cite{Guan12,Guan13,Guan14} we rely on persistence analysis for \lru as a foundation.
However, we go beyond their work in two ways:\looseness=-1
\begin{enumerate}
	\item We exploit both competitiveness \emph{and} block competitiveness, which allows to derive more precise bounds.%
	\item Our approach is applicable to microarchitectures that exhibit timing anomalies, while prior work had to assume their absence.
\end{enumerate}

We implemented our approach in the open-source WCET analysis tool \llvmta~\cite{Hahn22} and experimentally evaluate the analysis cost and precision on benchmarks from TACLeBench~\cite{Falk16}.
Our results show quantitative cache analysis for \fifo and \mru comes close to the precision of \lru and that analysis precision profits from combining competitiveness and block competitiveness.
At the same time, the cost of the analysis is similar to that of existing \lru analyses.

To summarize, in this paper, we make the following contributions:
\begin{itemize}
	\item We introduce the notion of block competitiveness.
	\item We systematically analyze the competitiveness and block competitiveness of \fifo and \mru relative to \lru.
	\item We show how to exploit competitiveness and block competitiveness in state-of-the-art WCET analysis without any assumptions about the presence of timing anomalies.
	\item We experimentally evaluate the benefits of quantitative cache analysis in terms of analysis precision and its cost in terms of analysis time.
\end{itemize}

\section{Preliminaries: Caches}\label{sec:preliminaries}

Caches are fast but small memories that buffer parts of main memory in order to bridge the speed gap between the processor and main memory.
Caches consist of \emph{cache lines}, which store data at the granularity of memory blocks $b \in \blocks$.
Memory blocks usually comprise a power-of-two number of bytes~$\BlockSize$, e.g. 64 bytes, so that the block $\block(a)$ that address $a$ maps to is determined by truncating the least significant bits of $a$, i.e., $\block(a) = \lfloor a/\BlockSize \rfloor$.
In order to facilitate an efficient cache lookup, caches are organized in \emph{sets} such that each memory block maps to a unique cache set~$\set(b) = b \bmod \NbSets$, where $\NbSets$ is the number of sets.
The number of cache lines~$\Associativity$ in each cache set is called the \emph{associativity} of the cache.

If an accessed block resides in the cache, the access \emph{hits} the cache.
Upon a cache \emph{miss}, the block is loaded from the next level of the hierarchy into its cache set.
If the cache set is full, another memory block from the same set has to be evicted to make room for the new block.
The block to evict is determined by the \emph{replacement policy}.

In this paper, we consider three popular replacement policies: \emph{least-recently-used} (\lru), \emph{first-in first-out} (\fifo), and \emph{not most-recently-used} (\mru).

\emph{Least-recently-used} (\lru) replaces the block that has been accessed least recently.
Thus, a memory block~$b$ hits in an \lru~cache of associativity~$\Associativity$ if $b$ has been accessed previously and less than $\Associativity$~distinct blocks in the same cache set have been accessed since the last access to~$b$.
LRU has been argued to be the most predictable replacement policy~\cite{Reineke2007}.

\emph{First-in, first-out} (\fifo) replaces blocks in the order in which they have entered the cache.
A memory block~$b$ hits in a \fifo~cache of associativity~$\Associativity$ if $b$ has been accessed previously and less than $\Associativity$ many misses have occurred in the same cache set since the last miss to~$b$.
Thus, unlike under \lru, hits to blocks in the cache do not affect future replacement decisions.

\fifo is known to be used in the \textsc{Intel XScale}, some \textsc{ARM9} and \textsc{ARM11}, and the \textsc{PowerPC75x} series.

\emph{Not most-recently-used} (\mru) maintains a \emph{use} bit for every cache line.
Whenever a block is accessed, its use bit is set to one.
The cache lines of every cache set are totally ordered from \emph{position} $0$ to $k-1$, where $k$ is associativity.
Upon a miss, the first cache line whose use bit is zero is used for replacement and its use bit is set to one.
Initially, the cache is empty and all use bits are set to zero.
Whenever a cache set's last use bit is set to one, a \emph{global bit flip} occurs, and all other bits are reset to zero.
In this way, there is always at least one cache line whose use bit is zero.
Furthermore, the most-recently-used block is always protected from replacement as its use bit is guaranteed to be set.

We illustrate the operation of \mru in \reffig{mru}.
The first access to block~$e$ is a miss, and block $a$ in position $0$ is replaced, as it is the first block with a use bit of zero.
The next access to block~$d$ is a hit.
Setting its use bit to one causes a global bit flip, resetting all other use bits to zero.
The following access to $a$ is a miss, and block $e$ in position $0$ is replaced again.
Thus, the next access to $e$ is a miss, and block $b$ in position $1$ is replaced.

\newcommand{\cachestate}[8]{[\underset{\tiny \text{#2}}{#1}\underset{\tiny \text{#4}}{#3}\underset{\tiny \text{#6}}{#5}\underset{\tiny \text{#8}}{#7}]}
\newcommand{\cachestatelong}[8]{[\underset{\tiny \text{#2}}{#1}\underset{\tiny \text{#4}}{#3}\underset{\tiny \text{#6}}{#5}\dots\underset{\tiny \text{#8}}{#7}]}

\begin{figure}
	\centering
	\begin{tikzpicture}[>=stealth, node distance=2cm, xscale=0.97] %
		\node (state1) at (0, 0) {$\cachestate{a}{0}{b}{1}{c}{1}{d}{0}$};
		\node (state2) at (2, 0) {$\cachestate{e}{1}{b}{1}{c}{1}{d}{0}$};
		\node (state3) at (4, 0) {$\cachestate{e}{0}{b}{0}{c}{0}{d}{1}$};
		\node (state4) at (6, 0) {$\cachestate{a}{1}{b}{0}{c}{0}{d}{1}$};
		\node (state5) at (8, 0) {$\cachestate{a}{1}{e}{1}{c}{0}{d}{1}$};

		\draw[->] (state1) -- node[above] {e} (state2);
		\draw[->] (state2) -- node[above] {d} (state3);
		\draw[->] (state3) -- node[above] {a} (state4);	  
		\draw[->] (state4) -- node[above] {e} (state5);	  
	\end{tikzpicture}
	\vspace{-6.5mm}
	\caption{Example of \mru operation.}\labelfig{mru}
	\vspace{-3.5mm}
\end{figure}

\mru is employed in various \textsc{Intel} microarchitectures including \textsc{Sandy Bridge}, \textsc{Westmere}, and \textsc{Nehalem}~\cite{Eklov11,Abel2020}.
It is also found in the \textsc{UltraSPARC T2}~\cite{Kongetira05}.
Note that in some of the prior work~\cite{Reineke07,Guan14}, \mru is referred to as \oldmru, in others~\cite{Malamy1994} it is referred to as \pseudomru, as it approximates \lru.
We use \mru in this paper as it is more descriptive of the policy's behavior than \oldmru and to avoid confusion with other policies that approximate \lru, such as tree-based \plru.

\section{Competitiveness and Block Competitiveness}\labelsec{definitions}

\subsection{Notation}
All replacement policies considered in this paper treat different cache sets independently of each other, \ie, memory accesses to a particular cache set do not affect the replacement decisions in other cache sets.
The behavior of a set-associative cache can thus be understood by considering each of its cache sets separately.
In the following, we model and analyze the behavior of fully-associative caches, which correspond to individual cache sets of a set-associative cache.

We denote the set of memory blocks as $\memoryBlocks$.
An \emph{access sequence} $\sigma$ is a finite sequence of memory blocks, \ie, $\sigma \in \memoryBlocks^*$.
The behavior of a replacement policy on an access sequence depends on state of the cache from which the sequence is processed.
We denote the set of cache states of policy $P$ as $\cacheSetStates{P}$ and the policy's initial, empty cache state as $\initialState{P}$.
Let $\update{P}(p,\sigma)$ denote the cache state after processing access sequence~$\sigma$ from cache state~$p$ under policy~$P$.

Further, $\misses{P}(p,\sigma)$ denotes the number of misses of policy~$P$ on access sequence~$\sigma$ and cache state~$p$.
Similarly, we denote the number of hits as $\hits{P}(p,\sigma)$.
In the example in \reffig{example}, $\sigma = \seq{a,b,c,c,b,d,b,e,b,f,f,b}$, $\misses{\fifo(2)}(\initialState{\fifo(2)},\sigma) = 8$ and $\hits{\fifo(2)}(\initialState{\fifo(2)},\sigma) = 4$.

Note that in this formal framework the associativity is captured as part of the policy as the number of hits and misses depends on it.
Thus \lru or \fifo are not policies but $\lru(5)$ and $\fifo(2)$ are, where the parameter denotes the associativity of the respective cache.

To focus on a specific memory block, we define $\misses{P,b}(p,\sigma)$ and $\hits{P,b}(p,\sigma)$ as the number of misses and hits of policy~$P$ on accesses to block $b$ in access sequence $\sigma$ and cache state~$p$.
In the example in \reffig{example}, $\misses{\lru(2),b}(\initialState{\lru(2)},\sigma) = 1$ and $\hits{\lru(2),b}(\initialState{\lru(2)},\sigma) = 4$.

\subsection{Hit and Miss Competitiveness}

We first recapitulate the definitions of \emph{miss competitiveness} and \emph{hit competitiveness} from Reineke and Grund~\cite{Reineke08}, which we adapt slightly as discussed below:
	
\index{relative competitiveness!definition}
\begin{restatable}[Miss Competitiveness~\cite{Reineke08}]{defi}{misscomp}\labeldef{misscompetitiveness}~\\
Policy $P$ is $(r,c)$-{miss-competitive} relative to policy $Q$ if \[\misses{P}(p,\sigma) \leq r \cdot \misses{Q}(\initialState{Q},\sigma) + c\] for all access 
sequences $\sigma \in \accessSequences$ and cache states $p \in \cacheSetStates{P}$.
\end{restatable}
\begin{restatable}[Hit Competitiveness~\cite{Reineke08}]{defi}{hitcompdefinition}\labeldef{hitcompetitiveness}~\\
Policy $P$ is $(r,c)$-{hit-competitive} relative to policy $Q$ if \[\hits{P}(p,\sigma) \geq r \cdot \hits{Q}(\initialState{Q},\sigma) - c\] for all access 
sequences $\sigma \in \accessSequences$ and cache states $p \in \cacheSetStates{P}$.
\end{restatable}

We will later instantiate the definitions above with $\fifo(k)$ and $\mru(k)$ as $P$ and $\lru(l)$ as $Q$ and use the relations to transfer guarantees obtained for \lru to \fifo and \mru.

The difference between the definitions above and those in~\cite{Reineke08} is that we restrict the initial state of $Q$ to be the empty initial state.
This is convenient for our purposes in this work as it 
\begin{itemize}
	\item[(a)] improves the additive part $c$ slightly, and
	\item[(b)] can still be exploited safely in WCET analysis as all guarantees derived for $\lru(l)$ cover the empty initial state.\looseness=-1
\end{itemize}

Previous work~\cite{Reineke08} has analyzed the relative competitiveness of \fifo and \plru relative to \lru, showing that $\lru(2k-1)$ is $(1,0)$-miss-competitive relative to $\fifo(k)$ and $\plru(k)$ is $(1,0)$-miss-competitive relative to $\lru(1+\log_2 k)$.
These results imply that existing classifying analyses for \lru can be transferred to \fifo and \plru.
While \cite{Reineke08} argues that relative competitiveness can also be used to transfer bounds on the number of hits and misses from \lru to other policies, it does  not discuss how to soundly incorporate this information within WCET analysis.

In \refsec{wcet}, we show how to exploit hit and miss competitiveness in WCET analysis for set-associative caches, applying the respective relations to each cache set separately.

\subsection{Block-Hit and Block-Miss Competitiveness}

We are now ready to introduce block-hit and block-miss competitiveness, which extend hit and miss competitiveness to focus on individual memory blocks.

\begin{restatable}[Block-Miss Competitiveness]{defi}{blockmisscomp}\labeldef{blockmisscompetitiveness}~\\
	Policy $P$ is $(r,c)$-{block-miss-competitive} relative to policy $Q$ if\looseness=-1
	\begin{equation}\label{eq:blockmiss}
		\misses{P,b}(p,\sigma) \leq r \cdot \misses{Q,b}(\initialState{Q},\sigma) + c
	\end{equation}
	for all access 
	sequences $\sigma \in \accessSequences$, memory blocks $b \in \memoryBlocks$, and cache states $p \in \cacheSetStates{P}$.
\end{restatable}

\begin{restatable}[Block-Hit Competitiveness]{defi}{blockhitcomp}\labeldef{blockhitcompetitiveness}~\\
	Policy $P$ is $(r,c)$-{block-hit-competitive} relative to policy $Q$ if\looseness=-1
	\begin{equation}
		\hits{P,b}(p,\sigma) \geq r \cdot \hits{Q,b}(\initialState{Q},\sigma) - c
	\end{equation}
	for all access 
	sequences $\sigma \in \accessSequences$, memory blocks $b \in \memoryBlocks$, and cache states $p \in \cacheSetStates{P}$.
\end{restatable}

Miss-block competitiveness allows to transfer persistence guarantees:
If block $b$ is persistent in $\lru(l)$ and $Q$ is $(r,c)$-block-miss-competitive relative to $\lru(l)$, then block $b$ can cause at most $r+c$ misses in $Q$.
If persistent blocks are accessed frequently this must result in many hits, which can be transferred via block-hit competitiveness.

We will see in more detail in \refsec{wcet} how to exploit block competitiveness in WCET analysis.

\subsection{General Properties of Competitiveness}

The following lemma is sometimes useful to obtain competitiveness statements from block competitiveness.
\begin{restatable}[Block competitiveness to Competitiveness]{lem}{blockhittohit}\labellem{blockhittohit}~ 

If policy $P$ is $(r,0)$-block-hit-competitive relative to policy $Q$, then policy $P$ is also $(r,0)$-hit-competitive relative to policy~$Q$.

	\spacedstyle{-20} If policy $P$ is $(r,0)$-block-miss-competitive relative to policy~$Q$, then policy $P$ is also $(r,0)$-miss-competitive relative to policy~$Q$.\spacedstyle{0}
\end{restatable}
\begin{proof}
	This follows directly from the fact that
	\begin{eqnarray*}
		\hits{P}(p, \sigma) &=& \sum_{b \in \memoryBlocks} \hits{P,b}(p, \sigma) \\
		\misses{P}(p, \sigma) &=& \sum_{b \in \memoryBlocks} \misses{P,b}(p, \sigma)
	\end{eqnarray*}
	\vspace{-5mm}~\\
\end{proof}
The converse does not hold in general.
\begin{restatable}[Equiv. of $(1,0)$-Hit and -Miss Competitiveness]{lem}{equivonehit}
	The following statements are equivalent:
	\begin{compactenum}
		\item \spacedstyle{-20} Policy $P$ is $(1,0)$-block-hit-competitive relative to policy $Q$.
		\item \spacedstyle{-25} Policy $P$ is $(1,0)$-block-miss-competitive relative to policy $Q$.
	\end{compactenum}
	The following statements are also equivalent:
	\begin{compactenum}{\setcounter{enumi}{2}}
		\item \spacedstyle{0}Policy $P$ is $(1,0)$-hit-competitive relative to policy $Q$.
		\item Policy $P$ is $(1,0)$-miss-competitive relative to policy $Q$.
	\end{compactenum}
\end{restatable}
The proof is omitted here for brevity and can be found in 
\ifthenelse{\isundefined\istechreport}{the technical report~\cite{Kahlen2025arxiv}}{the appendix}.

\subsection{Block Competitiveness for Set-Associative Caches}

Note that while we analyze block-hit and block-miss competitiveness for fully-associative caches, the results directly transfer to set-associative caches as we will now show.

Consider a set-associative cache with $\NbSets$ cache sets, where each cache set is independently controlled by policy~$P$.
Let $\misses{P,b, \NbSets}(p,\sigma)$ denote the number of misses to block~$b$ on access sequence~$\sigma$ from cache state $p = (p_0, \dots, p_{\NbSets-1}) \in C_P^\NbSets$, where $p_0, \dots, p_{\NbSets-1}$ are the cache states of the individual cache sets.
As each cache set is independent, the number of misses to block~$b$ only depends on the accesses in $\sigma$ that map to the same cache set as block~$b$:
\begin{equation}\label{eq:projection}
	\misses{P,b,\NbSets}(p,\sigma) = \misses{P,b}(\pi(p,\set(b)), \pi(\sigma,\set(b))),
\end{equation}
where $\pi$ projects cache states and access sequences to the corresponding cache set, \ie, $\pi(p,i) = p_i$, the state of cache set~$i$ in cache state $p$, and $\pi(\sigma,i)$ is the subsequence of $\sigma$ containing only those accesses that map to cache set~$i$.

If policy $P$ is $(r,c)$-block-miss-competitive relative to policy~$Q$, we thus have:
\begin{eqnarray*}
	\misses{P,b,\NbSets}(p,\sigma) & \overset{(\ref{eq:projection})}{=} & \misses{P,b}(\pi(p,\set(b)), \pi(\sigma,\set(b))) \\
	& \overset{(\ref{eq:blockmiss})}\leq & r \cdot \misses{Q,b}(\initialState{Q}, \pi(\sigma,\set(b))) + c \\
	&  \overset{(\ref{eq:projection})}{=} & r \cdot \misses{Q,b,\NbSets}(\initialState{Q}^\NbSets, \sigma) + c,
\end{eqnarray*}
for all access sequences $\sigma \in \accessSequences$, memory blocks $b \in \memoryBlocks$, and cache states $p \in \cacheSetStates{P}^\NbSets$.

Block-hit competitiveness can be transferred to set-associative caches analogously.

\section{Competitiveness of \fifo}\labelsec{fifo}

In this section we analyze the competitiveness of \fifo relative to \lru.
We start with a negative result:

\begin{restatable}[Block-Miss Competitiveness: $\fifo(k)$ vs $\lru(l)$]{theo}{blockmissfifolru}\labelthm{blockmissfifolru}
	For all $k \geq l \geq 2$, $\fifo(k)$ is not $(r,c)$-block-miss-competitive relative to $\lru(l)$ for any pair $(r, c)$.
	\end{restatable}
	\begin{proof}
	Let $\rho_i = \seq{b, a_1^i, a_2^i, \dots, a_{l-1}^i}$ where all $a_j^i$ are distinct from each other and from $b$.
	Let $\sigma_n = \rho_1 \circ \rho_2 \circ ... \rho_n$.
	Under $\lru(l)$ only the first access to $b$ in $\sigma_n$ misses.
	All subsequent accesses hit as exactly $l-1$ distinct blocks are accessed between any two consecutive accesses to $b$.
	Under $\fifo(k)$, on the other hand, the number of misses to $b$ grows with $n$.
	This is similar to the example we have seen in \reffig{example}.
	Thus, for any pair $(r, c)$ there is an $n$ such that \[\misses{\fifo(k),b}(\initialState{\fifo(k)},\sigma_n) > r \cdot \misses{\lru(l),b}(\initialState{\lru(l)},\sigma_n) + c.\]
	\end{proof}

	On the other hand, we can show that \fifo is miss competitive relative to \lru:
	\begin{restatable}[Miss Competitiveness: $\fifo(k)$ vs $\lru(l)$]{theo}{missfifolru}\labelthm{missfifolru}
	For all $k \geq l$, $\fifo(k)$ is $\left(\frac{k}{k-l+1}, 0\right)$-miss-competitive relative to $\lru(l)$.
	\end{restatable}
	\begin{proof}
	This follows immediately from the two following facts:
 	1. For $k \geq l$, $\fifo(k)$ is $\left(\frac{k}{k-l+1}, 0\right)$-miss-competitive relative to the optimal offline policy $\minAlg(l)$ %
	 for associativity~$l$~\cite[Theorem~6 and following discussion]{Sleator1985} starting from an empty initial state.
	2. $\minAlg(l)$ is optimal, \ie, it results in the fewest misses on every access sequence.
	In particular, $\lru(l)$ always exhibits at least as many misses as $\minAlg(l)$.
	\end{proof}
	
	$\fifo(k)$ is also block-hit-competitive relative to $\lru(l)$ and if $k$ is much larger than $l$, almost every $\lru(l)$ hit will also be a hit in $\fifo(k)$:
	\begin{restatable}[Block-Hit Competitiveness: $\fifo(k)$ vs $\lru(l)$]{theo}{blockhitfifolru}\labelthm{blockhitfifolru}
	
	For all $k \geq l$,  $\fifo(k)$ is 
	$\left(1-\frac{1}{\left\lceil\frac{k}{l-1}\right\rceil}, 0\right)$%
	-block-hit-competitive relative to $\lru(l)$.
	\end{restatable}
	\begin{proof}
	Let $\sigma \in \accessSequences$ be an arbitrary access sequence, $b \in \memoryBlocks$ be an arbitrary memory block, and $p \in \cacheSetStates{\fifo(k)}$ be an arbitrary state of $\fifo(k)$.
	We need to show that 
	\[\hits{\fifo(k),b}(p,\sigma) \geq \left(1-\frac{1}{\left\lceil\frac{k}{l-1}\right\rceil}\right) \cdot \hits{\lru(l),b}(\initialState{\lru(l)},\sigma).\]
	
	Assume $\sigma$ incurs $n$ misses to block $b$ in $\fifo(k)$, \ie, $\misses{\fifo(k),b}(p,\sigma) = n$.
	Then, we can partition $\sigma$ into $n+1$ subsequences $\sigma_1, \dots, \sigma_n, \sigma_{post}$, where each $\sigma_i$ ends with the access that causes the $i^{th}$ miss in $\fifo(k)$. $\sigma_{post}$ contains accesses following the final miss in $\sigma$, and may thus be empty.
	Let $h_{i, \fifo}$ and $h_{i, \lru}$ denote the number of hits to block~$b$ of $\fifo(k)$ and $\lru(l)$ in subsequence $\sigma_i$, respectively.
	Let $m_{i, \fifo}$ and $m_{i, \lru}$ be defined analogously for misses.
	By the definition of the subsequences $\sigma_i$, $m_{i, \fifo} = 1$ for $i \in \{1, \dots, n\}$.
	The number of accesses $a_i$ to block $b$ on $\sigma_i$ is $a_i = h_{i, \lru}+m_{i, \lru} = h_{i, \fifo}+m_{i, \fifo}$.

	As $\lru(l)$ starts in the empty initial state, the first access in $\sigma_1$ must be a miss.
	Thus $m_{1, \fifo} = 1 \leq m_{1, \lru}$ and so $h_{1, \fifo} \geq h_{1, \lru}$. 
	Also, $h_{post, \fifo} = a_{post} \geq h_{post, \lru}$.
	
	We will show that for all $i > 1$, $h_{i, \fifo} \geq \left(1-\frac{1}{\left\lceil\frac{k}{l-1}\right\rceil}\right) \cdot h_{i, \lru}$, which is sufficient to prove the theorem.
	
	On each subsequence $\sigma_i$ with $i > 1$, we can distinguish two cases: 1. $m_{i, \lru} \geq 1$ and 2. $m_{i, \lru} = 0$.
	In the first case, it follows immediately from $m_{i, \fifo} = 1$ that $h_{i, \fifo} \geq h_{i, \lru}$.
	Now, consider the second case, $m_{i, \lru} = 0$ and thus $h_{i, \fifo} = h_{i, \lru}-1$.
	As $\sigma_i$ and $\sigma_{i-1}$ end in misses to $b$ for $\fifo(k)$ the subsequence $\sigma_i$ must incur at least $k$ misses in $\fifo(k)$, otherwise the final access in $\sigma_i$ would yield a hit to $b$. Subsequence~$\sigma_i$ must thus include accesses to at least $k$ distinct memory blocks other than $b$.%
	On the other hand, between two consecutive accesses to block $b$ in $\lru(l)$, $\sigma_i$ must contain strictly less than $l$ distinct blocks for the second of the two accesses to be a hit in $\lru(l)$. %
	Thus, at most $l-1$ of the $k$~distinct blocks can be accessed before each hit to $b$ in $\lru(l)$.
	As a consequence, we get the following lower bound: $h_{i,\lru} \geq \left\lceil \frac{k}{l-1} \right\rceil$.

	As $h_{i, \fifo} = h_{i, \lru}-1$, $\frac{h_{i, \fifo}}{h_{i, \lru}} = \frac{h_{i, \lru}-1}{h_{i, \lru}} = 1-\frac{1}{h_{i, \lru}} \geq 1-\frac{1}{\left\lceil\frac{k}{l-1}\right\rceil}$, which concludes the proof.
	\end{proof}

	\refthm{blockhitfifolru} is strongly related to and inspired by \cite[Theorem 1]{Guan13}, where the authors derive an upper bound on the number of misses of \fifo on accesses to a block that is persistent in \lru in a loop.
	The exact same bound can be derived from \refthm{blockhitfifolru} for persistent blocks.
	\refthm{blockhitfifolru} is, however, slightly more general than the corresponding theorem in~\cite{Guan13} as it is not limited in applicability to loops and it can still be beneficial if only some accesses to a block in a loop are persistent.\looseness=-1

	Whenever \emph{all} blocks accessed in a loop collectively fit into the cache, Lemma 2 in \cite{Guan13} provides a better bound on the total number of misses.
	This property is not expressible using block-hit or block-miss competitiveness.
	However, bounds close the one provided by \cite[Lemma 2]{Guan13} can often be derived from \fifo's miss competitiveness.

	\begin{restatable}[Hit Competitiveness: $\fifo(k)$ vs $\lru(l)$]{cor}{hitfifolru}\label{cor:hitfifolru}~\\
		For all $k \geq l$,  $\fifo(k)$ is 
		$\left(1-\frac{1}{\left\lceil\frac{k}{l-1}\right\rceil}, 0\right)$
		-hit-competitive relative to $\lru(l)$.
	\end{restatable}
	This follows immediately from \refthm{blockhitfifolru} and \reflem{blockhittohit}.

\newcommand{\mruexample}{
\begin{figure*}[t]
	\centering
	\begin{tikzpicture}[>=stealth, node distance=2cm, xscale=1] %
		\node (state1) at (0, 0) {$\cachestate{u}{0}{v}{1}{w}{0}{x}{1}$};
		\node (state2) at (2, 0) {$\cachestate{\bf b}{1}{v}{1}{w}{0}{x}{1}$};
		\node (state3) at (4, 0) {$\cachestate{\bf b}{0}{v}{0}{w}{1}{x}{0}$};
		\node (state4) at (6, 0) {$\cachestate{u}{1}{v}{0}{w}{1}{x}{0}$};
		\node (state5) at (8, 0) {$\cachestate{u}{1}{\bf b}{1}{w}{1}{x}{0}$};

		\node (state6) at (10, 0) {$\cachestate{u}{0}{\bf b}{0}{w}{0}{x}{1}$};
		\node (state7) at (12, 0) {$\cachestate{u}{1}{\bf b}{0}{w}{0}{x}{1}$};
		\node (state8) at (14, 0) {$\cachestate{u}{1}{v}{1}{w}{0}{x}{1}$};
		\node (state9) at (16, 0) {$\cachestate{u}{0}{v}{0}{\bf b}{1}{x}{0}$};

		\draw[->] (state1) -- node[above] {\bf b} (state2);
		\draw[->] (state2) -- node[above] {w} (state3);
		\draw[->] (state3) -- node[above] {u} (state4);	  
		\draw[->] (state4) -- node[above] {\bf b} (state5);	  

		\draw[->] (state5) -- node[above] {x} (state6);
		\draw[->] (state6) -- node[above] {u} (state7);
		\draw[->] (state7) -- node[above] {v} (state8);
		\draw[->] (state8) -- node[above] {\bf b} (state9);

	\end{tikzpicture}
	\vspace{-2.5mm}
	\caption{Example of \mru operation in which block~$b$ is repeatedly accessed.}\labelfig{mruexample}
	\vspace{-3mm}
\end{figure*}}

\section{Competitiveness of \mru}\labelsec{nmru}

\mruexample

\begin{restatable}[Block-Hit and Block-Miss Competitiveness: $\mru(k)$ vs $\lru(2)$]{theo}{blockhitmisslrutwo}\labelthm{blockhitmisslrutwo}
	For all $k \geq 2$, $\mru(k)$ is $(1, 0)$-block-hit- and block-miss-competitive relative to $\lru(2)$.
\end{restatable}
\begin{proof}
	It suffices to show that $\mru(k)$ always contains the two most-recently-used blocks, once two blocks have been accessed.
	As these are the only blocks that can be in $\lru(2)$ states, any hit in $\lru(2)$ must also be a hit in $\mru(k)$ and the theorem follows.

	By construction, the \emph{use} bit of the most-recently-used block is always set in $\mru(k)$.
	Thus, the most-recently-used block cannot be replaced upon a miss. 
	As a consequence, the second-most-recently-used block must always be in the cache.
\end{proof}

\begin{restatable}[Hit and Miss Competitiveness: $\mru(k)$ vs $\lru(2)$]{cor}{hitmisslrutwo}\labelthm{hitmisslrutwo}
	For all $k \geq 2$, $\mru(k)$ is $(1, 0)$-hit- and miss-competitive relative to $\lru(2)$.
\end{restatable}
This follows immediately from \refthm{blockhitmisslrutwo} and \reflem{blockhittohit}.

\begin{restatable}[Block-Miss Competitiveness: $\mru(k)$ vs $\lru(l)$]{theo}{blockmissmrulru}\labelthm{blockmissmrulru}
	For all $k \geq l \geq 3$, $\mru(k)$ is $(l, 0)$-block-miss-competitive relative to $\lru(l)$.
	\end{restatable}
	\begin{proof}
	Let $\sigma \in \accessSequences$ be an arbitrary access sequence, $b \in \memoryBlocks$ be an arbitrary memory block, and $p \in \cacheSetStates{\mru(k)}$ be an arbitrary state of $\mru(k)$.
	We need to show that $\misses{\mru(k),b}(p,\sigma) \leq l \cdot \misses{\lru(l),b}(\initialState{\lru(l)},\sigma)$.
	
	Assume $\sigma$ incurs $n$ misses to block $b$ in $\lru(l)$, \ie, $\misses{\lru(l),b}(p,\sigma) = n$.
	Then, we can partition $\sigma$ into $n+1$ subsequences $\sigma_{pre}, \sigma_1, \dots, \sigma_n$, where each $\sigma_i$ begins with the access that causes the $i^{th}$ miss to $b$ and ends right before $\sigma_{i+1}$ begins or the entire sequence ends. $\sigma_{pre}$ contains the initial accesses in $\sigma$ before the first miss to $b$, and may thus be empty.
	
	As $\initialState{\lru(l)}$ does not contain $b$, the first access to $b$ in $\sigma$ must cause a miss.
	Thus, $\sigma_{pre}$ cannot contain any accesses to $b$.
	So there can also not be any misses to $b$ in $\sigma_{pre}$ in $\mru(k)$.
	
	We argue that at most $l$ misses may happen in each subsequence $\sigma_i$ in $\mru(k)$, which implies that $\misses{\mru(k),b}(p,\sigma) \leq l \cdot n = l\cdot \misses{\lru(l),b}(\initialState{\lru(l)},\sigma)$
	
	To get an intuition for why there can be at most $l$ misses in each subsequence $\sigma_i$ consider \reffig{mruexample}.
	Here, block $b$ is repeatedly accessed with up to three distinct blocks in between.
	After the first miss, $b$ is in position $0$.
	After the first global bit flip, $b$ gets replaced before it is accessed again.
	Thus the next access to $b$ must be a miss, but it will be inserted at a greater position than before, as all the bits to the left of $b$'s previous position have been set.
	Eventually, $b$ will be in position $l-1$ or greater, and no more misses to $b$ can happen, because $b$ must be accessed before the $l-1$ bits to its left could have been flipped, as no additional misses to $b$ occur in $\lru(l)$.

	Let $a_{i,j}$ be the position of $b$ in $\mru(k)$ after the $j^{th}$ miss to $b$ in $\sigma_i$.
	If $a_{i,j} \geq l-1$, then no more misses to $b$ are possible within $\sigma_i$:
	As no additional misses to $b$ occur for $\lru(l)$ within~$\sigma_i$, at most $l-1$ distinct other blocks are accessed between any two accesses to $b$.
	This implies in particular that at most $l-1$ distinct other blocks are accessed after each global bit flip (including the access that causes the global bit flip) before the next access to $b$.
	Each of these accesses flips at most one of the $l-1$ bits left of $b$. 
	Thus, $b$'s bit is flipped to $1$ before it can be replaced by a miss.
	
	Assume that $a_{i,j} < l-1$.
	Then, we argue below that $a_{i, j+1} \geq a_{i, j}+1$.
	As a consequence, $a_{i, j} \geq j-1$ and so $a_{i, l} \geq l-1$. Thus, at most $l$ misses may occur.

	For a $(j+1)^{st}$ miss to $b$ to occur, $b$ must be replaced following its insertion at position $a_{i, j}$ upon the $j^{th}$ miss.
	As $l \leq k$, there cannot be two global bit flip between two accesses to $b$.
	If $b$ has been evicted from position $a_{i, j}$, and no additional global bit flip has happened since $b$'s eviction, then the bits of all positions before must be set.
	Thus, $b$ will be inserted $a_{i, j+1} > a_{i, j}$.\looseness=-1
	\end{proof}

	\refthm{blockmissmrulru} is strongly related to and inspired by \cite[Theorem IV.3]{Guan12}, which bounds the number of misses of \mru on accesses to a block in a strongly-connected component of the CFG, where consecutive accesses to the block are separated by at most $l-1$ distinct blocks.
	Theorem IV.3 in \cite{Guan12} can be derived from \refthm{blockmissmrulru} and the fact that such a block is persistent in $\lru(l)$, splitting the argument into a workload-independent and a workload-dependent aspect.

	Sometimes the following miss-competitiveness result may provide better guarantees than \refthm{blockmissmrulru}:
	\begin{restatable}[Miss Competitiveness: $\mru(k)$ vs $\lru(l)$]{theo}{missmrulru}\labelthm{missmrulru}
	For all $k \geq l \geq 2$, $\mru(k)$ is $\left(\frac{k-1}{k-l+1}, l-2\right)$-miss-competitive relative to $\lru(l)$.
	\end{restatable}
\begin{proof}	
	First, note that at a global bit flip, $\mru(k)$ contains the $k$ most-recently accessed blocks:
	Between any two global bit flips in $\mru(k)$, exactly $k-1$ distinct blocks are accessed: 
	The first access to each of these distinct blocks increases the number of set bits by one; initially, one bit is set. 
	A global bit flip occurs when the $k$-th bit is set.

	Also, between any two consecutive global bit flips, no block that has been accessed after the first global bit flip can be evicted before the second global bit flip as its use bit remains set. 
	Thus, at a global bit flip, $\mru(k)$ contains all blocks that have been accessed since the last global bit flip.

	Let $\sigma \in \accessSequences$ be an arbitrary access sequence, $b \in \memoryBlocks$ be an arbitrary memory block, and $p \in \cacheSetStates{\mru(k)}$ be an arbitrary states of $\mru(k)$.
	We need to show that $\misses{\mru(k)}(p,\sigma) \leq \frac{k-1}{k-l+1} \cdot \misses{\lru(l)}(\initialState{\lru(l)},\sigma) + l-2$.
	
	We partition $\sigma$ along the global bit flips.
	Assume $\sigma$ results in $m$ global bit flips starting from $p$.
	Then we partition $\sigma$ into $m+1$ subsequences $\sigma_1, \dots, \sigma_m, \sigma_{post}$, where each $\sigma_i$ ends with the access that causes the $i^{th}$ global bit flip and $\sigma_{post}$ contains the accesses following the final global bit flip.
	By construction, $\sigma_{post}$ may be empty.

	Let $n_i$ be the number of misses to block $b$ in $\mru(k)$ in $\sigma_i$ and let $l_i$ be the number of misses to $b$ in $\lru(l)$ in $\sigma_i$.

	We will show that 
	\begin{enumerate}
		\item[a)] $n_i \leq \frac{k-1}{k-l+1}\cdot l_i$ for all $i$, and
		\item[b)] $n_{post} \leq l_{post} + l-2$.
	\end{enumerate}
	Then, the theorem follows from the fact that 
	\begin{align*}\small
	\misses{\mru(k)}(p, \sigma) &= \left(\sum_{i=1}^{m} n_i\right) + n_{post} \\ &\leq  \left(\sum_{i=1}^{m} \frac{k-1}{k-l+1} \cdot l_i\right) + l_{post} + l-2 \\
	&\leq \frac{k-1}{k-l+1} \cdot \left(\sum_{i=1}^{m} l_i + l_{post}\right) + l-2 \\ &= \frac{k-1}{k-l+1} \cdot \misses{\lru(l)}(\initialState{\lru(l)},\sigma) + l-2.
	\end{align*}
	
	a) Let us now show that $n_i \leq \frac{k-1}{k-l+1}\cdot l_i$ for all $i$.
	
	We first consider the case that $i > 1$.	
	Exactly $k-1$ distinct blocks get accessed in subsequence~$\sigma_i$ for $i > 1$.
	As each block accessed in a subsequence misses at most once in $\mru(k)$, the number of misses $n_i$ in $\mru(k)$ is at most $k-1$.
	
	At the global bit flip, $\lru(l)$ contains the most-recently-used block, which causes the global bit flip, and which is not part of these $k-1$ distinct blocks.
	Thus, at most $l-1$ of these blocks may hit on their first access in $\sigma_i$ in $\lru(l)$.
	Thus, at least $k-1-(l-1) = k-l$ misses will happen in $\lru(l)$.

	We distinguish two cases:
	\begin{enumerate}
		\item $l_i \geq k-l+1$.
		\item $l_i = k-l$.
	\end{enumerate}
	In the first case, $n_i \leq k-1 \leq \frac{k-1}{k-l+1} \cdot l_i$.

	In the second case, $\sigma_i$ must contains hits to all of the contents of $\lru(l)$ at the global bit flip. 
	In particular, it must hit the least-recently-used block. 
	Thus, no miss may occur in \lru before accessing the least-recently-used block. 
	As $\mru(k)$ contains all the blocks of $\lru(l)$ at the global bit flip, the access that hits the \lru block must also hit in \mru. 
	Thus, \mru may have at most $k-2$ misses and so $n_i \leq k-2 \leq \frac{k-2}{k-l}  \cdot l_i \leq \frac{k-1}{k-l+1} \cdot l_i$.

	We now consider the case that $i=1$.
	The first subsequence $\sigma_1$ is special as it starts with the empty initial state of $\lru(l)$.
	The first accesses to all blocks in $\sigma_i$ miss in $\lru(l)$.
	Each of these blocks misses at most once in $\mru(k)$.
	Thus, $n_1 \leq l_1$.

	b) Let us now show that $n_{post} \leq l_{post} + l-2$.
	   Let $d$ be the number of distinct blocks accessed in $\sigma_{post}$.
	   At most $d$ misses may happen in $\mru(k)$.
	   As argued above, at most $l-1$ of these blocks may hit in $\lru(l)$ and thus at least $d-(l-1)$ misses must happen in $\lru(l)$.
	   We again distinguish two cases:\looseness=-1
	   \begin{enumerate}
	   	\item $l_{post} \geq d-(l-1)+1 = d-l+2$.
	   	\item $l_{post} = d-(l-1)= d-l+1$.
	   \end{enumerate} 
	   In the first case, $n_{post} \leq d \leq l_{post} + l-2$.

	   In the second case, $\sigma_{post}$ must contain hits to all of the contents of $\lru(l)$ at the global bit flip.
	   By the same reasoning as above we must then have $n_{post} \leq d-1 \leq l_{post} + l-2$.
\end{proof}
	
	\begin{restatable}[Block-Hit Competitiveness: $\mru(k)$ vs $\lru(l)$]{theo}{blockhitmrulru}\labelthm{blockhitmrulru}
	For all $k \geq l \geq 3$, $\mru(k)$ is not $(r, c)$-block-hit-competitive relative to $\lru(l)$ for any $r > 0$ and any~$c$.%
	\end{restatable}
	\begin{proof}
		For any state $p \in \cacheSetStates{\mru(k)}$, any block~$b$, and any number of hits~$h$, we show below how to construct a sequence $\sigma_h$, such that $\hits{\mru(k),b}(p, \sigma_h) = 0$ and $\hits{\lru(l),b}(\initialState{\lru(l)}, \sigma_h) \geq h$.\looseness=-1

		Let $r > 0$ and $c$ be arbitrary.
		Pick $h > c/r$.
		Then $\hits{\mru(k),b}(p, \sigma_h) < r \cdot \hits{\lru(l),b}(\initialState{\lru(l)}, \sigma_h) - c$, which proves the theorem.

		Let $q \in \cacheSetStates{\mru(k)}$ be arbitrary.
		Below we show how to construct a sequence $\sigma(q)$, such that $\hits{\mru(k),b}(q, \sigma(q)) = 0$ and $\hits{\lru(l),b}(\initialState{\lru(l)}, \sigma(q)) = 1$.

		We can then use this construction to build $\sigma_h$ as follows:
		\begin{itemize}
			\item $\sigma_1 = \sigma(p)$.
			\item $\sigma_{n+1} = \sigma_n \circ \sigma(q)$, where $q = \update{\mru(k)}(p, \sigma_n)$.
		\end{itemize}

		Let us now show how to construct $\sigma(q)$.

		Let $\rho = \seq{a_1, a_2, a_3, \dots}$ be an infinite sequence of ``fresh'' memory blocks, \ie, blocks not contained in state~$q$ and distinct from each other and from $b$.
		
		Accessing any prefix of $\rho$ from state~$q$ will result in a sequence of misses, which will cause a sequence of global bit flips.
		Eventually, a global bit flip will happen at position $k-1$.
		Pick $\sigma_{pre}$ as the shortest prefix of $\rho$ that causes a global bit flip at position $k-1$.
		
		Let $q'$ be the result of accessing $\sigma_{pre}$ in state $q$, \ie, $q'=\update{\mru(k)}(q,\sigma_{pre}) = \cachestatelong{a_i}0{a_{i+1}}0{a_{i+2}}0{a_{i+k-1}}1$.
		
			Accessing $\sigma_{mid} = \seq{a_{i+2}, \dots, a_{i+k-2}}$ in $q'$ will flip all but the first two use bits:\\
			$q'' = \update{\mru(k)}(q', \sigma_{mid}) = \cachestatelong{a_i}0{a_{i+1}}0{a_{i+2}}1{a_{i+k-1}}1$.

			Accessing $\seq{b, a_{i+1}, a_{i}}$ in $q''$ results in a miss to $b$, a hit to $a_{i+1}$ that causes a global bit flip, and a miss to $a_i$ that replaces $b$.
		Let $\sigma(q) = \sigma_{pre} \cdot \sigma_{mid} \cdot \seq{b, a_{i+1}, a_{i}, b}$.
			By construction $\hits{\mru(k),b}(q, \sigma(q)) = 0$ and $\hits{\lru(l),b}(\initialState{\lru(l)}, \sigma(q)) = 1$ as the second access to $b$ is a hit in $\lru(l)$ for $l \geq 3$.
	\end{proof}

	While \refthm{blockhitmrulru} shows that $\mru(k)$ is not \emph{block}-hit-competitive relative to $\lru(l)$ for any $l \geq 3$, we can show that $\mru(k)$ is hit-competitive relative to $\lru(l)$ for $k \geq 2l$:
	
	\begin{restatable}[Hit Competitiveness: $\mru(k)$ vs $\lru(l)$]{theo}{hitmrulru}\labelthm{hitmrulru}
	For all $k \geq 2l$, $\mru(k)$ is $\left(1-\frac{1}{\lceil\frac{k}{2l}\rceil}, (1-\frac{1}{\lceil\frac{k}{2l}\rceil})\cdot(l-1)\right)$-hit-competitive relative to $\lru(l)$.%
	\end{restatable}
	\vspace{-1mm}

	\begin{table*}[ht!]
		\caption{Summary of competitiveness results for \fifo and \mru and the corresponding theorems.}\labeltab{summary}
		\centering
		\begin{tabular}{ccccc}
		\toprule
		Policy & miss competitiveness & block-miss competitiveness & hit competitiveness & block-hit competitiveness\\
		\midrule
		\fifo & \parbox{1.1in}{\centering$(\frac{k}{k-l+1},0)$\\ \refthms{missfifolru}, \cite[Th. 6]{Sleator1985}} & $\infty$ \refthms{blockmissfifolru}&
		$(1-\frac{1}{\lceil\frac{k}{l-1}\rceil},0)$ \refcors{hitfifolru} & $(1-\frac{1}{\lceil\frac{k}{l-1}\rceil},0)$ \refthms{blockhitfifolru} \\
		\mru & \parbox{1in}{\centering$(\frac{k-1}{k-l+1},l-2)$\\\centering \refthms{missmrulru}} &
		$\begin{cases}
			(1,0) & : l \leq 2 \text{ \refthms{blockhitmisslrutwo}}\\
			(l,0) & : k \geq l > 2 \text{ \refthms{blockmissmrulru}}
		\end{cases}$ 
		& \parbox{1.4in}{\small\centering$(1-\frac{1}{\lceil\frac{k}{2l}\rceil},(1-\frac{1}{\lceil\frac{k}{2l}\rceil})\cdot(l-1))$\\[1mm]\centering \normalsize\refthms{hitmrulru}}	& $\begin{cases}
			(1,0) & : l \leq 2 \text{ \refthms{blockhitmisslrutwo}}\\
			(0,-) & : l > 2 \text{ \refthms{blockhitmrulru}}
		\end{cases}$ \\
		\bottomrule
		\end{tabular}
		\end{table*}

Our strategy to prove \refthm{hitmrulru} is to partition a given access sequence along the global bit flips in $\mru(k)$.
We call each subsequence between two consecutive global bit flips a \emph{phase}, where bit flips are included in the phase that they start.
We start counting at $0$ and so the first phase is phase $0$.
However, unlike in the proof of \refthm{missmrulru}, we cannot simply relate the number of hits in $\mru(k)$ to the number of hits in $\lru(l)$ in the same phase.
Instead, we relate hits that are exclusive to $\lru(l)$ in phase $i+1$ to hits in $\mru(k)$ in phase $i$.\looseness=-1

\newcommand{\hitsBoth}[1]{L_{#1}}
\newcommand{\hitsLRU}[1]{\underline{\hitsBoth{#1}}}	
\newcommand{\hitsNMRU}[1]{\overline{\hitsBoth{#1}}}	
\newcommand{\blocksLRU}[1]{B(\hitsLRU{#1})}	
\newcommand{\blocksNMRU}[1]{B(\hitsNMRU{#1})}	
\newcommand{\nhitsBoth}[1]{n_{#1}}
\newcommand{\nhitsLRU}[1]{\underline{\nhitsBoth{#1}}}	
\newcommand{\nhitsNMRU}[1]{\overline{\nhitsBoth{#1}}}	

We distinguish three types of hits in a given phase:
\begin{enumerate}
	\item Hits under both policies. We denote the set of accesses that hit under both policies in phase $i$ as $\hitsBoth{i}$.
	\item Hits under $\lru(l)$ but not under $\mru(k)$, which we denote by $\hitsLRU{i}$.
	\item Hits under $\mru(k)$ but not under $\lru(l)$, which we denote by $\hitsNMRU{i}$.
\end{enumerate}
This is illustrated by the following Venn diagram:
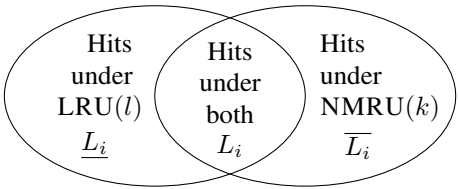
\begin{center}
	\begin{tikzpicture}[yscale=0.6]
	
		\draw (0,0) circle (2cm);
		\draw (2,0) circle (2cm);
	
		\node[text width=1cm,align=right] at (-0.8, 0.4) {Hits under $\lru(l)$};
		\node[text width=1cm,align=center] at (1, 0.3) {Hits under both};
		\node[text width=1cm] at (2.7, 0.4) {Hits under $\mru(k)$};
	
		\node at (-0.8, -1.1) {$\hitsLRU{i}$};
		\node at (1, -1.1) {$\hitsBoth{i}$};
		\node at (2.7, -1.1) {$\hitsNMRU{i}$};
	\end{tikzpicture}
\end{center}

Given a set of memory blocks $B$, let $\hitsBoth{i}(B)$ be the set of accesses to blocks in $B$ that hit under both policies in phase~$i$.
We similarly define the restrictions of $\hitsLRU{i}$ and $\hitsNMRU{i}$ to $B$ as $\hitsLRU{i}(B)$ and $\hitsNMRU{i}(B)$, respectively.
The set of memory blocks accessed by the accesses in $\hitsLRU{i}$ and $\hitsNMRU{i}$ is denoted as $\blocksLRU{i}$ and $\blocksNMRU{i}$, respectively.

We denote the number of hits under both policies to blocks in $B$ in phase~$i$ as $\nhitsBoth{i}(B) = |\hitsBoth{i}(B)|$, the number of hits under $\lru(l)$ but not under $\mru(k)$ as $\nhitsLRU{i}(B) = |\hitsLRU{i}(B)|$, and the number of hits under $\mru(k)$ but not under $\lru(l)$ as $\nhitsNMRU{i}(B) = |\hitsNMRU{i}(B)|$.

The following lemma shows that if we can bound the number of exclusive LRU hits in phase $i+1$ in a certain way by the number of \mru hits in phase $i$, then this implies hit competitiveness of \mru relative to \lru.

\begin{restatable}[Hit Competitiveness: $\mru(k)$ vs $\lru(l)$]{lem}{hitcomp}\labellem{hitmrulru}
	Let~$\sigma$ be an arbitrary access sequence, $p \in \cacheSetStates{\mru(k)}$ be an arbitrary state of $\mru(k)$, and let $\nhitsBoth{i}(B), \nhitsLRU{i}(B), \nhitsNMRU{i}(B)$ be defined as above for any set of blocks $B$. 
	
	If
	\begin{align}
		\forall i \geq 1: \exists B_i, \overline{B_i}: \notag\\
		B_i \cup \overline{B_i} &= \blocksLRU{i+1}~\wedge\label{boundzero}\\
		\nhitsNMRU{i} &\geq \nhitsLRU{i+1}(B_i)~\wedge\label{boundone}\\
		\nhitsBoth{i} + \nhitsNMRU{i} &\geq x\cdot\nhitsLRU{i+1}(\overline{B_i})\label{boundtwo}
	\end{align}
	then
	\begin{eqnarray}
		\hits{\mru(k)}(p,\sigma) \geq \notag\\&\hspace{-1cm} \left(1-\frac{1}{x+1}\right)\cdot\left(\hits{\lru(l)}(\initialState{\lru(l)},\sigma) - (l-1)\right)\notag
	\end{eqnarray}
\end{restatable}
The proof of this lemma and the proof of \refthm{hitmrulru} can be found in \ifthenelse{\isundefined\istechreport}{the technical report~\cite{Kahlen2025arxiv}}{the appendix}.

\reftab{summary} summarizes the competitiveness results for \fifo and \mru obtained in this paper.

\section{Exploiting Competitiveness in WCET Analysis}\label{sec:wcet}

\subsection{WCET Analysis in a Nutshell}
We integrate quantitative cache analysis into \llvmta~\cite{Hahn22} an open-source WCET analysis tool that implements the \emph{de facto} standard architecture of static WCET analysis tools~\cite{Wilhelm08}.

\llvmta operates on a control-flow-graph (CFG) representation of the program under analysis.
The CFG is not explicit in the machine code executed on the hardware.
Thus most WCET analysis tools rely on the reconstruction of the CFG from the binary.
\llvmta, instead, is integrated into the \textsc{llvm} compiler infrastructure~\cite{Lattner04} and thus obtains the CFG during compilation.\looseness=-1

\llvmta performs three analysis steps:
\begin{enumerate}
	\item \emph{Control-flow analysis} derives constraints on the possible control flow of the program. In particular it determines bounds on the number of times each loop may be executed.
	If the analysis is not able to determine a loop bound, the user needs to manually annotate a bound.
	\item \emph{Microarchitectural analysis} determines the possible execution times of parts of the program, such as basic blocks, accounting for all features of the underlying microarchitecture that may affect the execution time.
	\item \emph{Implicit path enumeration} (IPET)~\cite{Li1995} combines the results of the control-flow and microarchitectural analysis to determine the WCET.
\end{enumerate}
In \refsec{basicipet}, we recapitulate a basic implicit path enumeration technique.
In \refsec{exploitingcompetitiveness}, we then show how to exploit competitiveness results on top of this basic technique.

\subsection{Microarchitectural Analysis in a Nutshell}
Microarchitectural analysis within \llvmta can be seen as a cycle-by-cycle simulation of the program's execution on abstract hardware states.
The abstract hardware states capture information about the state of the cache, the pipeline, and the other features that may affect timing.
However, due to abstraction, this simulation may lack information required to determine the precise successor state, \eg, when a memory access cannot be classified as a guaranteed cache hit or miss.
In such cases, the simulation branches, creating multiple successor states.
While this branching may lead to a large number of states, it allows the analysis to safely bound the execution time of the program even in the presence of timing anomalies~\cite{Lundqvist1999}.

The output of the microarchitectural analysis is an \emph{abstract execution graph}~\cite{Stein2010} $G = (V, E, I, T)$ whose set of nodes $V$ are abstract hardware states and whose edges are transitions between these states.
Here, $I \subseteq V$ the set of initial nodes, $T \subseteq V$ is the set of terminal nodes, and $E \subseteq V \times V$ the set of edges.
Each edge is associated with a cost~$c_e$ corresponding to its number of execution cycles.

In addition, edges are labeled with microarchitectural events such as cache hits or misses, the memory block being accessed, and with program events, \eg, entering a loop or taking a back edge of a loop.
This information may be used to constrain the number of times particular edges can be taken during the implicit path enumeration step.

\subsection{Scope-Based Persistence Analysis in a Nutshell}

Persistence analysis is performed as part of the microarchitectural analysis.
It determines \emph{persistent edges} in the abstract execution graph that correspond to accesses that may only miss once during the program execution or during a certain scope of the program.
Some accesses are persistent in the entire program.
However, many accesses are only persistent within a certain \emph{scope} of the program.
A miss edge that is persistent within a scope may be taken once each time the control flow enters the scope.

In our persistence analysis, scopes correspond to the loops in the program.
Nested loops consequently result in nested scopes, inducing a partial order on scopes.

In our implementation, we employ the conditional-must persistence analysis~\cite{Reineke2018b} to determine which accesses are persistent accesses under \lru replacement for each associativity.
As \lru is a stack algorithm~\cite{Mattson1970}, a persistence analysis for associativity~$k$ is sufficient to determine the persistence of accesses for all associativities $a \leq k$.

\subsection{Basic IPET}\labelsec{basicipet}

IPET implicitly enumerates all possible paths through the abstract execution graph $G$ via a set of integer linear constraints and reduces the problem of determining the longest path to an integer linear programming (ILP) problem.
To this end, each edge in the graph is associated with an integer variable~$x_e$ that captures how often the edge is traversed.

The objective is to maximize the execution time, which is expressed as follows:
\begin{equation*}
\max \sum_{e \in E} c_e \cdot x_e
\end{equation*}
Exactly one of the edges leaving one of the initial nodes~$I$ must be taken:
\begin{equation*}
\sum_{e = (v,w) \in E, v \in I} x_e = 1
\end{equation*}
For each non-extremal node of the graph, the in- and the out-flow must be equal:
\begin{equation*}
	\forall w \in V \setminus (I \cup T): \sum_{e = (v,w) \in E} x_e = \sum_{e = (w,v) \in E} x_e
\end{equation*}
\newcommand{\backedge}{\textit{BE}}
\newcommand{\loopbound}{\textit{LB}}
\newcommand{\entryedge}{\textit{EE}}
\newcommand{\edge}{\textit{E}}
\newcommand{\blockaccesses}{\textit{A}}
\newcommand{\blockmisses}{\textit{M}}
\newcommand{\blockhits}{\textit{H}}

Let $\backedge_l$ be the set of back edges of loop $l$, $\entryedge_l$ be the set of entry edges of loop $l$, and $\loopbound_l$ be the loop bound of loop~$l$.
The following constraints capture the loop bounds:
\begin{equation*}
\forall l \in L: \sum_{e \in \backedge_l} x_l \leq \loopbound_l \cdot \sum_{e \in \entryedge_l} x_e
\end{equation*}

\subsection{Exploiting Persistence via Competitiveness within IPET}\labelsec{exploitingcompetitiveness}

At the high level, we apply the following strategy to exploit persistence under \lru in the analysis of caches employing other replacement policies:
\begin{enumerate}
	\item We introduce integer variables that capture the \emph{hypothetical} number of cache misses under \lru for each block and associativity and constrain these variables using the persistence analysis results and the \emph{sequence of memory accesses} encoded by the variables~$x_e$.
	\item We bound the number of misses on the \emph{microarchitectural path}, also encoded by the variables~$x_e$, based on the hypothetical number of misses under \lru and the competitiveness results.
\end{enumerate}
This renders some microarchitectural paths, encoded by the variables~$x_e$, infeasible--specifically, those that would result in a higher number of misses than permitted by the number of misses under \lru and the competitiveness guarantees.

For each scope $s$, let $\entryedge_s$ be the set of entry edges of scope~$s$ and let $\edge_s$ denote the set of all edges belonging to scope~$s$.

For each block $b$, let $\blockaccesses_b$ denote the set of edges with accesses to block~$b$.
Similarly, let $\blockmisses_b$ and $\blockhits_b$ denote the set of edges with misses and hits to block~$b$, respectively, \ie, $\blockhits_b = \blockaccesses_b \setminus \blockmisses_b$.
We can constrain these sets to the edges in a certain scope by intersecting them with the set of edges in the scope:
\begin{equation*}
	\blockaccesses_{b,s} = \blockaccesses_b \cap \edge_s, \quad \blockmisses_{b,s} = \blockmisses_b \cap \edge_s, \quad \blockhits_{b,s} = \blockhits_b \cap \edge_s
\end{equation*}

\newcommand{\persistentaccesses}{\textit{P}}
\newcommand{\nonpersistentaccesses}{\textit{NP}}

For each block $b$, each scope $s$, and each associativity $a$, let $\persistentaccesses_{b,s,a}$ be the set of edges with accesses to $b$ that are classified as persistent in scope $s$ for associativity $a$.

\newcommand{\bestscope}[1]{\textit{BS}(#1)}

Under \lru, all accesses to a block $b$ that are persistent within scope~$s$ may result in at most one miss each time the control flow enters scope $s$.
However, the same block may have accesses that are persistent in multiple scopes. 
The nesting structure of scopes induces a partial order on scopes:
a scope~$s$ is \emph{outer} to a scope $s'$ if $s$ is a proper ancestor of $s'$.

\newcommand{\essentialscope}{\textit{ES}}

The outermost scope in which an access is persistent induces the best bound on the number of misses.
In order to cover all persistent accesses, we determine a block's set of \emph{essential scopes} $\essentialscope_{b, a}$ for every associativity~$a$.
A scope is essential for block $b$ and associativity $a$ if there is at least one access to $b$ that is persistent in this scope and no outer scope:
\begin{equation*}
	\essentialscope_{b, a} = \{ s \mid \persistentaccesses_{b,s,a} \neq \emptyset \land \forall s' \text{ outer to } s:  \persistentaccesses_{b,s',a} \not\supseteq \persistentaccesses_{b,s,a} \}
\end{equation*}

\newcommand{\lrumisses}[1]{\textit{m}_{#1,\lru}}
\newcommand{\lruhits}[1]{\textit{h}_{#1,\lru}}
\newcommand{\Pmisses}[1]{\textit{m}_{#1}}
\newcommand{\Phits}[1]{\textit{h}_{#1}}
\newcommand{\accesses}[1]{\textit{a}_{#1}}

For each block $b$ and associativity $a$, we introduce an integer variable~$\lrumisses{b,a}$ that captures the number of misses to block~$b$ under \lru~with associativity $a$.
We bound this number of misses as follows:
\begin{equation}\label{eq:lrublockmisses}
\lrumisses{b,a} \leq \sum_{s \in \essentialscope_{b,a}, e \in \entryedge_s} x_e + \sum_{e \in \blockaccesses_b \setminus \bigcup_{s \in \essentialscope_{b,a}} \persistentaccesses_{b,s,a}} x_e
\end{equation}
The first summand upper bounds the number of persistent misses using the block's essential scopes\footnote{This approach is conservative in that it may account for misses to blocks that are not accessed on the microarchitectural path. Experimenting with a more accurate formulation is future work.}, while the second summand counts all other accesses\footnote{It is necessary to consider edges from both $\blockhits_b$ and $\blockmisses_b$, as we want to bound the hypothetical number of misses of \lru.}.

For each block $b$ we introduce variables $\Pmisses{b}, \Phits{b}$, and $\accesses{b}$ that capture the number of misses, the number of hits, and the number of accesses to block $b$, respectively, using the following constraints:

\begin{equation*}
   \Pmisses{b} = \sum_{e \in \blockmisses_{b}} x_e, \quad
   \Phits{b} = \sum_{e \in \blockhits_{b}} x_e, \quad
   \accesses{b} = \sum_{e \in \blockaccesses_{b}} x_e
\end{equation*}

To bound the number of misses to block $b$ we add a constraint for every associativity~$a$, exploiting the block-miss competitiveness of the policy under analysis relative to \lru:
\begin{equation}\label{eq:blockmisscomp}
   \Pmisses{b} \leq r(a) \cdot \lrumisses{b,a} + c(a)
\end{equation}
where the policy under analysis is $(r(a),c(a))$-block-miss-competitive relative to \lru~with associativity $a$.

To exploit block-hit competitiveness, we need to first determine a lower bound on the number of hits to each block $b$ under \lru.
To this end, we introduce a variable $\lruhits{b,a}$ for each block $b$ and associativity $a$ and constrain it as follows:
\begin{equation*}
   \lruhits{b,a} \geq \accesses{b} - \lrumisses{b,a}
\end{equation*}

We then add a constraint for each block $b$ and associativity~$a$ exploiting the block-hit competitiveness relative to \lru:
\begin{equation}\label{eq:blockhitcomp}
   \Phits{b} \geq r(a) \cdot \lruhits{b,a} - c(a)
\end{equation}
where the policy under analysis is $(r(a),c(a))$-block-hit-competitive relative to \lru~with associativity $a$.

In order to exploit non-block competitiveness, we need to determine upper and lower bounds on the hypothetical number of hits and misses under \lru in each cache set, which we obtain by summing up the corresponding bounds across all blocks mapping to the same cache set:
\begin{eqnarray*}
	\lruhits{s, a} = \sum_{b \in \blocks, \set(b) = s} \lruhits{b,a}\\
	\lrumisses{s, a} = \sum_{b \in \blocks, \set(b) = s} \lrumisses{b,a}
\end{eqnarray*}

Similarly, we determine the number of hits and misses under the policy under analysis in each cache set:
\begin{equation*}
	\Phits{s} = \sum_{b \in \blocks, \set(b) = s} \Phits{b}, \quad
	\Pmisses{s} = \sum_{b \in \blocks, \set(b) = s} \Pmisses{b}
\end{equation*}

Finally, we add constraints exploiting the miss and hit competitiveness of the policy under analysis relative to \lru for each cache set $s$ and associativity $a$:
\begin{eqnarray}
   \Pmisses{s} \leq r(a) \cdot \lrumisses{s, a} + c(a)\label{eq:misscomp}\\
   \Phits{s} \geq r(a) \cdot \lruhits{s, a} - c(a)\label{eq:hitcomp}
\end{eqnarray}

\section{Related Work}
Static cache analysis is a key component of WCET analysis for modern processors.
Lv et al.~\cite{Lv2016} provide a survey of the state-of-the-art in this area, which can broadly be classified into two categories: \emph{classifying} analyses~\cite{Alt1996,Grund2009,Grund2010, Grund2010b, Chattopadhyay2013,Touzeau2017,Touzeau2019}, which guarantee that individual memory accesses hit or miss, and \emph{quantitative} analyses~\cite{Guan12,Guan13,Guan14}, usually based on persistence analysis~\cite{Arnold94,Mueller94,White97,Ferdinand97,Ferdinand1999rts,Mueller2000rts,Ballabriga2008,Cullmann2011,Huynh2011,Nagar2012,Nagar2012thesis,Cullmann2013tecs,Cullmann2013thesis,Zhang15,Stock2019,Reineke2018b,Brandner2022}, which provide bounds on the number of cache hits and misses.

This paper unifies the two most closely related lines of work on non-\lru policies by Guan, Lv, Yang, Yu, and Yi~\cite{Guan12,Guan13,Guan14} and Reineke and Grund~\cite{Reineke08}. %

Our work explains most aspects of existing quantitative analyses~\cite{Guan12,Guan13,Guan14} as the combination of (a) general, workload-independent \emph{block competitiveness} properties of the underlying policies relative to \lru, and (b) workload-dependent properties captured by \lru \emph{persistence analysis}.
We generalize the applicability of these properties in the context of WCET analysis to microarchitectures that may exhibit timing anomalies.
In \refsec{experiments}, we evaluate the effect of combining competitiveness and block competitiveness on the accuracy and cost of quantitative cache analysis.

Previous work on relative competitiveness~\cite{Reineke08} focussed on the derivation of classifying analyses.
This work is the first to carry out its application to quantitative cache analysis.\looseness=-1

\section{Experimental Evaluation}\labelsec{experiments}
The main aim of our experiments is to evaluate the
accuracy and cost of quantitative cache analysis for \fifo and \mru compared with a state-of-the-art \lru cache analysis.
In addition, we want to analyze whether block competitiveness and plain competitiveness complement each other to obtain better WCET bounds.

\begin{figure*}[ht!]
\centering
	\input{rtas_graphs/WCET_for_selected_programs.pgf}
	\vspace{-5mm}
	\caption{WCET bounds for different competitiveness-based analyses for \mru ($k = 4$, $\textit{size} = 4$~KB) normalized to the \emph{\lru may/must+persistence analysis}.\label{fig:ex_block_vs_block_miss_nmru}}
	\vspace{-1mm}
\end{figure*}

\begin{figure*}[ht!]
	\centering
	\input{rtas_graphs/WCET_for_selected_programs_fifo.pgf}
	\vspace{-5mm}
	\caption{WCET bounds for different competitiveness-based analyses for \fifo ($k = 4$, $\textit{size} = 4$~KB) normalized to the \emph{\lru may/must+persistence analysis}.\label{fig:ex_block_vs_block_miss_fifo}}
	\vspace{-1mm}
\end{figure*}

\subsection{Experimental Setup}
For all our experiments we assume a single-core five-stage in-order Arm processor with the following characteristics:
\begin{itemize}
	\item Independent instruction und data cache that are of the same size, use the same associativity and use 16-byte cache lines. 
	\item A unified main memory with a fixed access latency of 10 cycles for the first word of a request and then one word per cycle for the other words of the same request.%
	\item The caches use write-through no-write-allocate as write policy.
\end{itemize}

We used the following benchmarks from TACLeBench~\cite{Falk16}: \textsl{adpcm\_dec, adpcm\_enc, audiobeam, basicmath, binarysearch, bsort, cjpeg\_wrbmp, complex\_updates, countnegative, cover, epic, fft, filterbank, fir2dim, fmref, g723\_enc, gsm\_dec, gsm\_encode, h264\_dec, huff\_dec, iir, insertsort, jfdctint, lift, lms, ludcmp, matrix1, md5, minver, ndes, petrinet, pm, powerwindow, prime, rijndael\_dec, rijndael\_enc, sha, st, statemate, susan, test3}.\\
We had to exclude benchmarks that \llvmta was not able to analyze, in particular those involving recursion, which is currently not supported.
All benchmarks were compiled to \textsc{llvm} using \textsc{clang} with following options: \textsl{-w -S -gline-tables-only -O0 -Xclang -disable-O0-optnone -fno-builtin -target arm -march=armv4t -mfloat-abi=hard -emit-llvm}, 
then linked and then post processed using \textsl{opt -S unoptimized.ll -mem2reg -indvars -loop-simplify -instcombine -globaldce -dce}.

We configured \llvmta to use Gurobi as the MILP solver.
Each experiment was run on one logical core of an AMD Ryzen Threadripper PRO 5995WX processor. 

We implemented the following analyses for \lru:
\begin{itemize}
	\item \emph{\lru may/must}: Classical may- and must-analysis~\cite{Ferdinand1999rts} without persistence analysis.
	\item \emph{\lru may/must+persistence}: Classical may- and must-analysis \emph{and} conditional-must persistence analysis~\cite{Reineke2018b}.
\end{itemize}
To determine the limits of any analysis, we consider two extremal cases:
\begin{itemize}
	\item \emph{all hit}: Assumes that every access is a hit.
	\item \emph{all miss}: Assumes that every access is a miss.
\end{itemize}
As an additional baseline we consider two analyses that are trivially correct for any replacement policy as the most-recently-used block is cached under any replacement policy:
\begin{itemize}
	\item \emph{direct-mapped must}: Classical must-analysis~\cite{Ferdinand1999rts} for associativity one.
	\item \emph{direct-mapped must+persistence}: Classical must-analysis~\cite{Ferdinand1999rts} \emph{and} conditional-must persistence analysis~\cite{Reineke2018b} for associativity one.
\end{itemize}
For both \fifo and \mru we consider the following five analysis configurations.
\begin{itemize}
	\item \emph{block-miss competitiveness}: Uses only block-miss competitiveness~(\ref{eq:blockmisscomp}).
	\item \emph{block-hit competitiveness}: Uses only block-hit competitiveness~(\ref{eq:blockhitcomp}).
	\item \emph{miss competitiveness}: Uses only miss competitiveness~(\ref{eq:misscomp}).
	\item \emph{hit competitiveness}: Uses only hit competitiveness~(\ref{eq:hitcomp}).
	\item \emph{all combined comp.}: Combination of all of the above.
\end{itemize}
These five analyses additionally include the \emph{direct-mapped must+persistence} analysis, which is correct for any replacement policy.
The \emph{must} analysis significantly reduces the size of the abstract execution graph and thus the analysis cost.\looseness=-1

\begin{figure}[ht!]
	\centering
	\input{rtas_graphs/Misses_for_different_assotiativities.pgf}
	\vspace{-5mm}
	\caption{Maximal number of misses for the best analyses for \lru, \mru, and \fifo for different associativities and 64 cache sets. The results are normalized to the \lru \emph{must/may+persistence analysis} for associativity $k = 1$ and summarized across all benchmarks via the geometric mean.\label{fig:acc_over_k}}
	\vspace{-1mm}
\end{figure}
\begin{figure}[ht!]
	\centering
	\input{rtas_graphs/WCET_with_varying_size_k=4.pgf}
	\vspace{-5mm}
	\caption{WCET bounds for the best analyses for \lru, \mru, and \fifo for different cache sizes at a fixed associativity of $k=4$ normalized to \lru \emph{must/may+persistence analysis} for size $0.5$~KB and summarized across all benchmarks via the geometric mean.\label{fig:acc_over_s}}
	\vspace{-1mm}
\end{figure}

\subsection{Analysis Accuracy}

\reffig{ex_block_vs_block_miss_nmru} and \reffig{ex_block_vs_block_miss_fifo} show the WCET bounds obtained for \mru and \fifo, respectively, for all analyzed benchmarks, normalized to the WCET bound obtained by the \emph{\lru may/must+persistence analysis}.
We omitted the \emph{all hit} and \emph{all miss} cases here, as including them would make the other cases hard to distinguish.
For \fifo, we additionally omit \emph{block-miss competitiveness} as it is vacuous for \fifo due to \refthm{blockmissfifolru}.

For both \mru and \fifo we observe that the analysis that combines all competitiveness constraints comes quite close to the \emph{\lru may/must+persistence analysis}.
For \mru there are benchmarks on which \emph{miss competitiveness} outperforms \emph{block-miss competitiveness} (\eg, \textsl{iir}), and vice versa (\eg, \textsl{gsm\_dec}).
Thus, combining the two is beneficial.
\emph{Hit} and \emph{block-hit competitiveness} are clearly less effective than \emph{miss} and \emph{block-miss competitiveness} for \mru.

For \fifo, \emph{miss competitiveness} is the most effective individual competitiveness-based analysis.
However, we note there are benchmarks on which \emph{block-hit competitiveness} can improve the WCET bound slightly (\eg, \textsl{binarysearch}).

While the \emph{direct-mapped must+persistence} analysis is effective on some benchmarks, there are many benchmarks on which it is clearly outperformed by the competitiveness-based analyses, showing that the analyses are able to effectively exploit associativities greater than one.

In some cases, the bound on the number of misses of \lru obtained by (\ref{eq:lrublockmisses}) can be highly inaccurate, which in turn results in inaccurate WCET bounds for the competitiveness-based analyses. 
The prime example of this phenomenon is the \textsl{cover} benchmark.
This benchmark consists of a large switch statement in a loop.
The switch statement as a whole does not fit into the cache, but each individual case of the switch statement does.
Thus, all blocks are persistent within the scope of the switch statement, but not persistent in the surrounding loop.
However, following (\ref{eq:lrublockmisses}) the analysis pessimistically accounts for one miss per persistent block upon entering the scope of the switch statement.
In contrast, IPET for \lru persistence only accounts for misses for accesses that actually lie on the considered microarchitectural path.
It is future work to improve the analysis to avoid such cases.

In \reffig{acc_over_k} we analyze the effect of increasing the cache associativity on the predicted worst-case number of misses for the best analysis of \lru, \mru and \fifo, respectively, while keeping the number of cache sets fixed at 64.
We are thus considering caches of size $1$~KB to $8$~KB. 
The results across all benchmarks are summarized by computing the geometric mean of the predicted number of misses relative to the \emph{\lru must/may + persistence} analysis for $k = 1$.
Our main observation is that additional associativity and thus cache capacity is a lot more beneficial for data caches than for instruction caches.
A possible explanation is that the TACLeBench benchmarks are too small to stress larger instruction caches.
Also, a greater share of the improvement in the miss bound carries over to \fifo and \mru for data caches.

In \reffig{acc_over_s} we analyze the effect of increasing the cache size while keeping the associativity fixed at $4$ on the WCET bounds.
We consider cache sizes from $0.5$~KB to $16$~KB, corresponding to $8$ to $256$ cache sets.
The results across all benchmarks are summarized via the geometric mean of the WCET bounds relative to the \emph{\lru must/may + persistence} analysis for the smallest considered cache size. 
Unsurprisingly, increasing the cache size yields better WCET bounds under all analyses.
However, the relative gains are more pronounced for the competitiveness-based analyses than for \lru.
This may be explained by the fact that the benchmarks mostly fit into $1$~KB \lru caches, and so further increasing the cache size does not help under \lru.
In contrast, the competitiveness-based analyses profit from the additional cache capacity due to improved competitiveness factors relative to smaller \lru caches.\looseness=-1

\subsection{Analysis Cost}
\begin{figure}[ht!]
	\centering
	\input{rtas_graphs/Runtime_with_varying_size_k=4.pgf}
	\vspace{-6mm}
	\caption{Cost of different analysis for different cache sizes (all with $k = 4$) normalized to \lru \emph{must/may+persistence analysis} for size $0.5$~KB and summarized across all benchmarks via the geometric mean.\label{fig:runtime_over_s}}
	\vspace{-1mm}
\end{figure}
\begin{figure}[ht!]
	\centering
	\input{rtas_graphs/cost_for_different_analyses.pgf}
	\vspace{-6mm}
	\caption{Cost of different analyses for a cache with $k = 4$ and $\textit{size} = 4$~KB normalized to the LRU may/must + persistence analysis and summarized across all benchmarks via the geometric mean.\looseness=-1\label{fig:cost_diff_analysis}}
	\vspace{-1mm}
\end{figure}

\reffig{runtime_over_s} shows the geometric mean of the analysis runtimes across all benchmarks for varying cache sizes relative to the runtime of the \emph{\lru may/must+persistence analysis} for cache size $0.5$~KB.
Unsurprisingly, analysis runtimes increase with the cache size.
To our surprise, the \emph{competitiveness-based analyses} are often faster than the \emph{\lru may/must+persistence analysis}.
This is explained by the additional may analysis and the more complex must analysis performed by the \lru analysis, which in turn do not offer a significant advantage in terms of accuracy.

\reffig{cost_diff_analysis} shows the runtime and peak memory usage of the different analyses relative to the \emph{\lru may/must+persistence analysis}; again we determine the geometric mean across all benchmarks.
The competitiveness-based analyses do require more memory than the \emph{\lru may/must+persistence analysis}. 
On average they require around 20\% to 30\% more memory as seen in \reffig{cost_diff_analysis}.
Interestingly, combining all constraints influences memory usage only marginally and the runtime is sometimes even reduced.
Apparently more constrained models can be easier to optimize.

More detailed experimental results across all benchmarks can be found in \ifthenelse{\isundefined\istechreport}{the technical report~\cite{Kahlen2025arxiv}}{the appendix}.

\section{Conclusions}

In this work we have unified two existing lines of work towards cache analysis for non-LRU policies:
1) Reineke and Grund's {relative competitiveness} approach, and 2) Guan, Lv, Yang, Yu, and Yi's work on quantitative cache analysis for \fifo and \mru.
Most of the existing work on cache analysis for non-LRU policies can thus be explained as special cases of our framework.
Further, we have shown that the combination of competitiveness \emph{and} block competitiveness can lead to better WCET bounds than either of the two alone.

In future work, we would like to explore automating the analysis of block competitiveness to reduce the reliance of manual analysis and to obtain more accurate results.

\section{Acknowledgments}

We would like to thank the anonymous reviewers for their valuable comments and suggestions.
This work has received funding from the European Research Council under the European Union’s Horizon 2020 research and innovation programme (grant agreement No. 101020415).

\bibliographystyle{IEEEtran}
\balance
\bibliography{blockcomp}

\begin{techreport}
\newpage
\onecolumn

\appendices

\section{Proofs}

\equivonehit*
\begin{proof}
	To show that 1) and 2) are equivalent, note that
	the number of accesses~$\maccesses{P,b}(p, \sigma)$ to a block $b$ is the sum of the number of hits and misses to $b$:
	$\maccesses{P,b}(p, \sigma) = \hits{P,b}(p, \sigma) + \misses{P,b}(p, \sigma)$.
	Also, the number of accesses to $b$ is the same for $P$ and $Q$ and independent of the initial state:
	$\maccesses{P,b}(p, \sigma) = \maccesses{Q,b}(\initialState{Q}, \sigma)$.
	Thus, 1) implies 2) as
	\begin{eqnarray*}
		\misses{P,b}(p, \sigma) &=& \maccesses{P,b}(p, \sigma) - \hits{P,b}(p, \sigma) \\
		&=& \maccesses{Q,b}(\initialState{Q}, \sigma) - \hits{P,b}(p, \sigma) \\
		&\overset{1)}{\leq}& \maccesses{Q,b}(\initialState{Q}, \sigma) - \hits{Q,b}(\initialState{Q}, \sigma) \\
		&=& \misses{Q,b}(\initialState{Q}, \sigma).
	\end{eqnarray*}
	2) implies 1) analogously:
	\begin{eqnarray*}
		\hits{P,b}(p, \sigma) &=& \maccesses{P,b}(p, \sigma) - \misses{P,b}(p, \sigma) \\
		&=& \maccesses{Q,b}(\initialState{Q}, \sigma) - \misses{P,b}(p, \sigma) \\
		&\overset{2)}{\geq}& \maccesses{Q,b}(\initialState{Q}, \sigma) - \misses{Q,b}(\initialState{Q}, \sigma) \\
		&=& \hits{Q,b}(\initialState{Q}, \sigma).
	\end{eqnarray*}
	3) and 4) can be shown to be equivalent analogously.

\end{proof}

\hitcomp*
\begin{proof}
\begin{eqnarray*}
	\hits{\lru(l)}(\initialState{\lru(l)},\sigma)  &=&  \left(\sum_{i \geq 0} \nhitsBoth{i} + \nhitsLRU{i}\right) \\
	&=& \nhitsBoth{0} + \nhitsLRU{0} + \nhitsBoth{1} + \nhitsLRU{1} + \left(\sum_{i \geq 2} \nhitsBoth{i} + \nhitsLRU{i}\right) \\
	&\overset{\nhitsLRU{0} = 0}{=}& \nhitsBoth{0} + \nhitsBoth{1} + \nhitsLRU{1} + \left(\sum_{i \geq 2} \nhitsBoth{i} + \nhitsLRU{i}\right) \\
	&\overset{\nhitsLRU{1} \leq l-1}{\leq}& \nhitsBoth{0} + \nhitsBoth{1} + (l-1) + \left(\sum_{i \geq 2} \nhitsBoth{i} + \nhitsLRU{i}\right) \\
	&=& \nhitsBoth{0} + \nhitsBoth{1} + (l-1) + \left(\sum_{i \geq 2} \nhitsBoth{i} + \nhitsLRU{i}(B_{i-1}) + \nhitsLRU{i}(\overline{B_{i-1}})\right) \\
	&\overset{(\ref{boundone})+(\ref{boundtwo})}{\leq}& \nhitsBoth{0} + \nhitsBoth{1} + (l-1) + \left(\sum_{i \geq 2} \nhitsBoth{i} + \nhitsNMRU{i-1} + \frac{1}{x}(\nhitsBoth{i-1} + \nhitsNMRU{i-1})\right) \\
	&=& \left(\sum_{i \geq 0} \nhitsBoth{i}\right) + (l-1) + \left(\sum_{i \geq 1} \nhitsNMRU{i}\right) + \left(\frac{1}{x}\sum_{i \geq 1}\nhitsBoth{i} + \nhitsNMRU{i}\right) \\
	& \leq&\left(\sum_{i \geq 0} \nhitsBoth{i}\right) + (l-1) + \left(\sum_{i \geq 1} \nhitsNMRU{i}\right) + \left(\frac{1}{x}\sum_{i \geq 1}\nhitsBoth{i}+\nhitsNMRU{i}\right) \\
	&\leq& \left(\sum_{i \geq 0} \nhitsBoth{i} + \nhitsNMRU{i}\right) + \frac{1}{x}\left(\sum_{i \geq 0}\nhitsBoth{i} + \nhitsNMRU{i}\right) + (l-1)\\	&=& \left(1+\frac{1}{x}\right)\cdot\hits{\mru(k)}(p,\sigma) + (l-1)
\end{eqnarray*}
Thus we have
\begin{eqnarray*}
	\hits{\lru(l)}(\initialState{\lru(l)},\sigma) &\leq& \left(1+\frac{1}{x}\right)\cdot\hits{\mru(k)}(p,\sigma) + (l-1)\\
	\Leftrightarrow \frac{x+1}{x}\cdot\hits{\mru(k)}(p,\sigma) + (l-1) &\geq& \hits{\lru(l)}(\initialState{\lru(l)},\sigma)\\
	\Leftrightarrow \hits{\mru(k)}(p,\sigma) + (l-1) &\geq& \frac{x}{x+1}\left(\hits{\lru(l)}(\initialState{\lru(l)},\sigma) - (l-1)\right)\\
	\Leftrightarrow \hits{\mru(k)}(p,\sigma) + (l-1) &\geq& \left(1-\frac{1}{x+1}\right)\left(\hits{\lru(l)}(\initialState{\lru(l)},\sigma) - (l-1)\right)\\
\end{eqnarray*}
\end{proof}

Let us now prove \refthm{hitmrulru} using \reflem{hitmrulru}.
\hitmrulru*
\begin{proof}
Let $\sigma$ be an arbitrary access sequence, $p \in \cacheSetStates{\mru(k)}$ be an arbitrary state of $\mru(k)$.
To prove the theorem we apply \reflem{hitmrulru} with $x = \left\lceil\frac{k}{2l}\right\rceil-1$.
To apply \reflem{hitmrulru}, we need to find sets $B_i, \overline{B_i}$ that satisfy (\ref{boundzero}), (\ref{boundone}), and (\ref{boundtwo}).

Let $B_i = \blocksLRU{i+1} \cap \blocksNMRU{i}$ and $\overline{B_i} = \blocksLRU{i+1} \setminus \blocksNMRU{i}$.

By construction $B_i \cup \overline{B_i} = \blocksLRU{i+1}$ and so (\ref{boundzero}) is satisfied. 

Next we show that $B_i$ satisfies (\ref{boundone}) of \reflem{hitmrulru}:
Note that under \mru, a block can only cause one miss in a phase.
Thus any block may contribute at most one hit to the exclusive LRU hits in any phase.
On the other hand, every block in $B_i$ must contribute at least one exclusive $\mru$ hit in phase $i$.
Thus, $\nhitsLRU{i+1}(B_i) = |B_i| \leq \nhitsNMRU{i}(B_i) \leq \nhitsNMRU{i}$.

Now let us show that $\overline{B_i}$ satisfies (\ref{boundtwo}) of \reflem{hitmrulru} for $x=\lceil\frac{k}{2l}\rceil-1$.

\newcommand{\xphasei}{0.5}
\newcommand{\xphaseii}{4.5}
\newcommand{\xphaseiii}{8.5}

\begin{figure}
\begin{centering}
\begin{tikzpicture}
	\fill[red!20, opacity=0.5] (0, -0.8) rectangle (\xphasei, 1.1);
	\fill[blue!20, opacity=0.5] (\xphasei, -0.8) rectangle (\xphaseii, 1.1);
	\fill[green!20, opacity=0.5] (\xphaseii, -0.8) rectangle (\xphaseiii, 1.1);

	\draw[dashed, red] (\xphasei, -0.8) -- (\xphasei, 1.1) node[above, xshift=6.5mm,yshift=-5mm] {Phase $i$};
	\draw[dashed, blue] (\xphaseii, -0.8) -- (\xphaseii, 1.1) node[above, xshift=9.5mm,yshift=-5mm] {Phase $i+1$};

	\draw[->] (0, 0) -- (\xphaseiii, 0);
	
	\foreach \x in {2.5,8,6.3} {
		\draw (\x, 0.2) -- (\x, -0.2); %
	}
	
	\node[below] at (2.5, -0.1) {\strut$x$};
	\node[below] at (7.8, -0.1) {\strut$x$ miss};
	\node[below] at (6.3, -0.1) {\strut$x$ evicted};

	\node[above] at (1.5, 0) {$\sigma_{i, pre}$};
	\node[above] at (3.5, 0) {$\sigma_{i, post}$};
	\node[above] at (5.4, 0) {$\sigma_{i+1, pre_1}$};
	\node[above] at (7.15, 0) {$\sigma_{i+1, pre_2}$};
	
	\end{tikzpicture}
	\caption{Partition of phase $i$ and phase $i+1$.}\labelfig{mruphases}
\end{centering}
\end{figure}

Let $a$ be the minimum number of accesses to blocks in $\overline{B_i}$ in phase $i$.
As all but the first of these accesses must be hits in \mru, there must be $a-1$ hits for each block in $\overline{B_i}$.

Let $x$ be a block in $\overline{B_i}$ that is accessed exactly $a$ times in phase $i$.
Let $p$ be $x$'s position after the first access to $x$ in phase~$i$.
As illustrated in \reffig{mruphases}, we partition phase $i$ into two subsequences $\sigma_{i, pre}$ and $\sigma_{i, post}$, where $\sigma_{i, pre}$ contains ends before the first access to $x$ in phase $i$ and $\sigma_{i, post}$ contains the remaining accesses of the phase.
We further partition the accesses prior to the first access to $x$ in phase $i+1$ into $\sigma_{i+1, pre_1}$ and $\sigma_{i+1, pre_2}$, where $\sigma_{i+1, pre_1}$ contains the accesses in phase $i+1$ up to and including the access that evicts $x$ from the \mru cache.\looseness=-1

By assumption $x$ is accessed $a$ times in phase $i$.
Further, it cannot be accessed in phase $i+1$ before it is evicted, otherwise it would be protected from eviction in phase~$i+1$.
Thus, there are $a-1$ accesses to $x$ in $\sigma_{i, post}\circ \sigma_{i+1, pre_1}$.
As the first access to $x$ hits in \lru in phase $i+1$, at most $l-1$ distinct blocks may be accessed between any two accesses to $x$.
Thus, we can conclude that $|B(\sigma_{i, post}\circ \sigma_{i+1, pre_1}) \setminus \{x\}| \leq (a-1)\cdot(l-1)$, where we use $B(\rho)$ to refer to the set of blocks accessed in sequence $\rho$.

\newcommand{\dotleft}{.\textit{left}}
\newcommand{\dotright}{.\textit{right}}

Given a set of blocks~$B$ accessed in phase~$i$ or $i+1$, we refer to the set of blocks that occupy positions less than or equal to $x$'s position by $B\dotleft$ and to the set of blocks that occupy positions greater than $x$'s by $B\dotright$.

Note that $B(\sigma_{i, post})\dotright \cap B(\sigma_{i+1, pre_1})\dotleft = \emptyset$.
This is because before $x$'s eviction in phase $i+1$, all blocks to its right are still cached in phase $i+1$ and still to its right.
Also, $B(\sigma_{i, post})\dotright \cup B(\sigma_{i+1, pre_1})\dotleft \subseteq B(\sigma_{i, post}\circ \sigma_{i+1, pre_1}) \setminus \{x\}$.

There are $p$ positions strictly smaller than $x$'s.
Thus, $B(\sigma_{i+1, pre_1})\dotleft \geq p + 1$, as $\sigma_{i+1, pre_1}$ evicts $x$.
So $|B(\sigma_{i, post})\dotright| \leq (a-1)\cdot(l-1) - (p + 1)$

There are $k-p-1$ positions strictly larger than $x$'s.
Thus we also have $|B(\sigma_{i, pre})\dotright \cup B(\sigma_{i, post})\dotright| \geq k - p - 1$.

Combining the two above inequalities, we get
\begin{eqnarray*}
	|B(\sigma_{i, pre})\dotright| & \geq & k - p - 1 - |B(\sigma_{i, post})\dotright|\\
								 & \geq & k - p - 1 - (a-1)\cdot(l-1) + (p + 1)\\
								 & = & k - (a-1)\cdot(l-1)
\end{eqnarray*}

$B(\sigma_{i, pre})\dotright$ may contain the block that started phase $i$ via a global bit flip.
Except for this access, all accesses in $\sigma_{i, pre}$ to the right of $x$ must be cache hits as they are to the right of the first access to $x$ but occurred before the first access to~$x$.
Thus $|B(\sigma_{i, pre})\dotright| - 1 \geq k -1 - (a-1)\cdot(l-1)$ is a lower bound on the number of hits in $\mru(k)$ in phase $i$.

To summarize, we have two lower bounds on the number of hits of $\mru(k)$ in phase $i$:
\begin{enumerate}
	\item $lb_1(a) = (a-1)\cdot \nhitsLRU{i+1}(\overline{B_i})$, and
	\item $lb_2(a) = k - 1 - (a-1)\cdot(l-1)$.
\end{enumerate}
Both lower bounds are a function of $a$, the minimum number of accesses to blocks in $\overline{B_i}$ in phase $i$.
As $a$ is unknown, we need to determine a lower bound on the maximum of the two lower bounds $lb_1(a)$ and $lb_2(a)$: $\min_{a} \max \{lb_1(a), lb_2(a)\}$.

Note that $lb_1(a)$ is increasing in $a$ and $lb_2(a)$ is decreasing in $a$, as illustrated below:
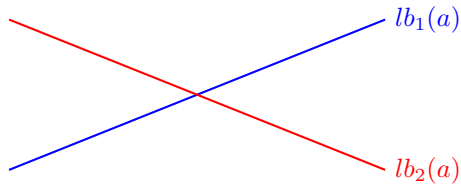
\begin{center}
\begin{tikzpicture}
    
    \draw[thick, blue] (0, 0.5) -- (5, 2.5) node[anchor=south west,yshift=-3mm] {$lb_1(a)$};
    
    \draw[thick, red] (0, 2.5) -- (5, 0.5) node[anchor=north west,yshift=3mm] {$lb_2(a)$};
\end{tikzpicture}
\end{center}

Thus for any value of $a$, the minimum of $lb_1(a)$ and $lb_2(a)$ is less than or equal to $\min_{a} \max \{lb_1(a), lb_2(a)\}$.

We pick $a = \left\lceil\frac{k}{2l}\right\rceil$ and show that both $lb_1(a)$ and $lb_2(a)$ are greater than or equal to $(\lceil\frac{k}{2l}\rceil-1)\cdot \nhitsLRU{i+1}(\overline{B_i})$.

$lb_1(a) = \left(\left\lceil\frac{k}{2l}\right\rceil-1\right)\cdot \nhitsLRU{i+1}(\overline{B_i}) \geq (\lceil\frac{k}{2l}\rceil-1)\cdot \nhitsLRU{i+1}(\overline{B_i})$.

$lb_2(a) = k - 1 - \left(\left\lceil\frac{k}{2l}\right\rceil-1\right)\cdot(l-1) \geq k - 1 - \frac{k}{2l}\cdot(l-1) = \frac{k}{2} -1 + \frac{k}{2l} \geq \frac{k}{2} = \frac{k}{2l}\cdot l \geq \frac{k}{2l}\cdot \nhitsLRU{i+1}(\overline{B_i}) \geq (\lceil\frac{k}{2l}\rceil-1)\cdot\nhitsLRU{i+1}(\overline{B_i})$.

So $\nhitsBoth{i} + \nhitsNMRU{i} \geq \min\{lb_1(a), lb_2(a)\} \geq (\lceil\frac{k}{2l}\rceil-1)\cdot\nhitsLRU{i+1}(\overline{B_i})$.
\end{proof}

\newpage
\section{Additional Experimental Results}
 \input{./rtas_graphs_all/Data_Cache_Misses_with_varying_associativity_64_cache_sets.pgf}
 \input{./rtas_graphs_all/Instruction_Cache_Misses_with_varying_associativity_64_cache_sets.pgf}
 \input{./rtas_graphs_all/WCET_with_varying_associativity_64_cache_sets.pgf}
 \input{./rtas_graphs_all/Runtime_with_varying_associativity_64_cache_sets.pgf}
 \input{./rtas_graphs_all/Memory_usage_with_varying_associativity_64_cache_sets.pgf}
 \input{./rtas_graphs_all/Data_Cache_Misses_with_varying_size_k=4_sub.pgf}
 \input{./rtas_graphs_all/Instruction_Cache_Misses_with_varying_size_k=4_sub.pgf}
 \input{./rtas_graphs_all/Runtime_with_varying_size_k=4_sub.pgf}
 \input{./rtas_graphs_all/Memory_usage_with_varying_size_k=4_sub.pgf}
 \input{./rtas_graphs_all/Data_Cache_Misses_k=4_size=1.0KB.pgf}
 \input{./rtas_graphs_all/Instruction_Cache_Misses_k=4_size=1.0KB.pgf}
 \input{./rtas_graphs_all/WCET_k=4_size=1.0KB.pgf}
 \input{./rtas_graphs_all/Runtime_k=4_size=1.0KB.pgf}
 \input{./rtas_graphs_all/Memory_Usage_k=4_size=1.0KB.pgf}
 \input{./rtas_graphs_all/Data_Cache_Misses_k=4_size=4.0KB.pgf}
 \input{./rtas_graphs_all/Instruction_Cache_Misses_k=4_size=4.0KB.pgf}
 \input{./rtas_graphs_all/WCET_k=4_size=4.0KB.pgf}
 \input{./rtas_graphs_all/Runtime_k=4_size=4.0KB.pgf}
 \input{./rtas_graphs_all/Memory_Usage_k=4_size=4.0KB.pgf}
 \input{./rtas_graphs_all/Data_Cache_Misses_k=4_size=2.0KB.pgf}
 \input{./rtas_graphs_all/Instruction_Cache_Misses_k=4_size=2.0KB.pgf}
 \input{./rtas_graphs_all/WCET_k=4_size=2.0KB.pgf}
 \input{./rtas_graphs_all/Runtime_k=4_size=2.0KB.pgf}
 \input{./rtas_graphs_all/Memory_Usage_k=4_size=2.0KB.pgf}
 \input{./rtas_graphs_all/Data_Cache_Misses_k=4_size=8.0KB.pgf}
 \input{./rtas_graphs_all/Instruction_Cache_Misses_k=4_size=8.0KB.pgf}
 \input{./rtas_graphs_all/WCET_k=4_size=8.0KB.pgf}
 \input{./rtas_graphs_all/Runtime_k=4_size=8.0KB.pgf}
 \input{./rtas_graphs_all/Memory_Usage_k=4_size=8.0KB.pgf}
 \input{./rtas_graphs_all/Data_Cache_Misses_k=2_size=2.0KB.pgf}
 \input{./rtas_graphs_all/Instruction_Cache_Misses_k=2_size=2.0KB.pgf}
 \input{./rtas_graphs_all/WCET_k=2_size=2.0KB.pgf}
 \input{./rtas_graphs_all/Runtime_k=2_size=2.0KB.pgf}
 \input{./rtas_graphs_all/Memory_Usage_k=2_size=2.0KB.pgf}
 \input{./rtas_graphs_all/Data_Cache_Misses_k=4_size=16.0KB.pgf}
 \input{./rtas_graphs_all/Instruction_Cache_Misses_k=4_size=16.0KB.pgf}
 \input{./rtas_graphs_all/WCET_k=4_size=16.0KB.pgf}
 \input{./rtas_graphs_all/Runtime_k=4_size=16.0KB.pgf}
 \input{./rtas_graphs_all/Memory_Usage_k=4_size=16.0KB.pgf}
 \input{./rtas_graphs_all/Data_Cache_Misses_k=8_size=8.0KB.pgf}
 \input{./rtas_graphs_all/Instruction_Cache_Misses_k=8_size=8.0KB.pgf}
 \input{./rtas_graphs_all/WCET_k=8_size=8.0KB.pgf}
 \input{./rtas_graphs_all/Runtime_k=8_size=8.0KB.pgf}
 \input{./rtas_graphs_all/Memory_Usage_k=8_size=8.0KB.pgf}
 \input{./rtas_graphs_all/Data_Cache_Misses_k=4_size=0.5KB.pgf}
 \input{./rtas_graphs_all/Instruction_Cache_Misses_k=4_size=0.5KB.pgf}
 \input{./rtas_graphs_all/WCET_k=4_size=0.5KB.pgf}
 \input{./rtas_graphs_all/Runtime_k=4_size=0.5KB.pgf}
 \input{./rtas_graphs_all/Memory_Usage_k=4_size=0.5KB.pgf}
 \input{./rtas_graphs_all/Data_Cache_Misses_k=1_size=1.0KB.pgf}
 \input{./rtas_graphs_all/Instruction_Cache_Misses_k=1_size=1.0KB.pgf}
 \input{./rtas_graphs_all/WCET_k=1_size=1.0KB.pgf}
 \input{./rtas_graphs_all/Runtime_k=1_size=1.0KB.pgf}
 \input{./rtas_graphs_all/Memory_Usage_k=1_size=1.0KB.pgf}

\end{techreport}

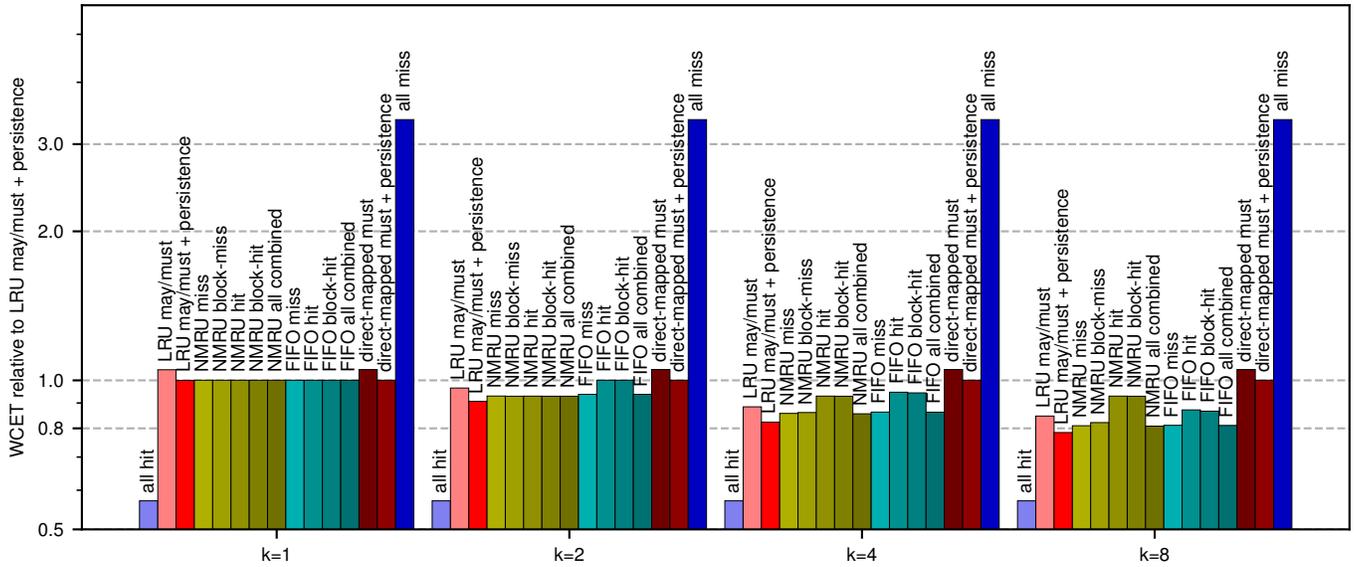
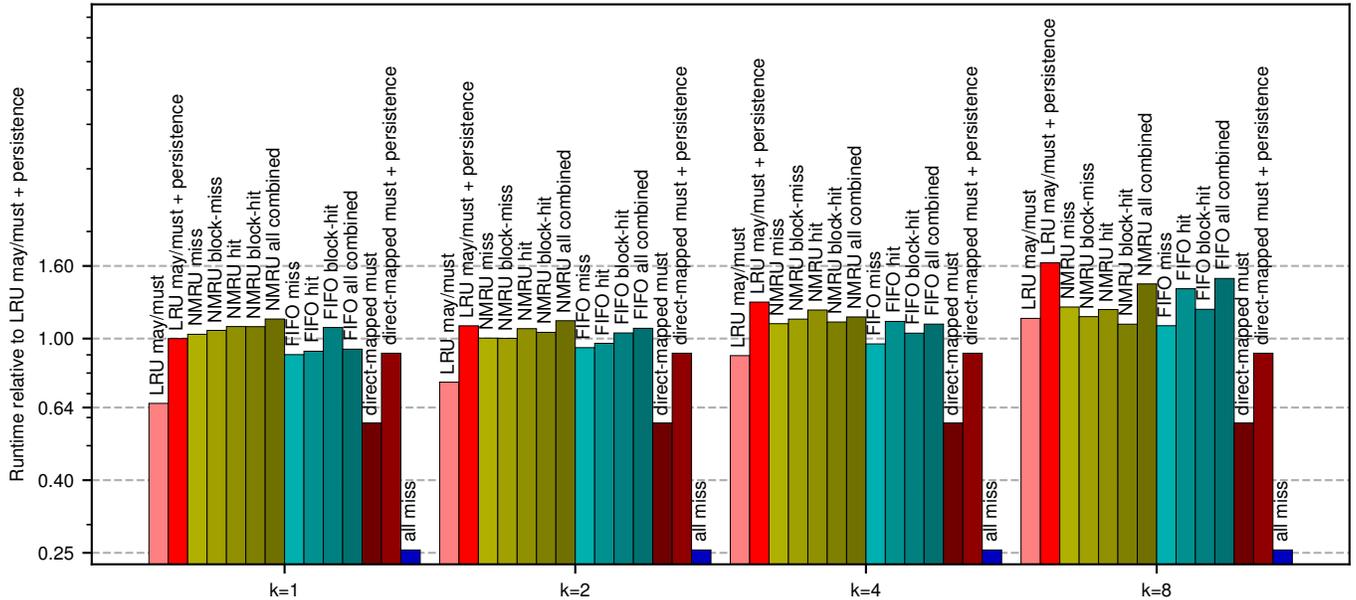
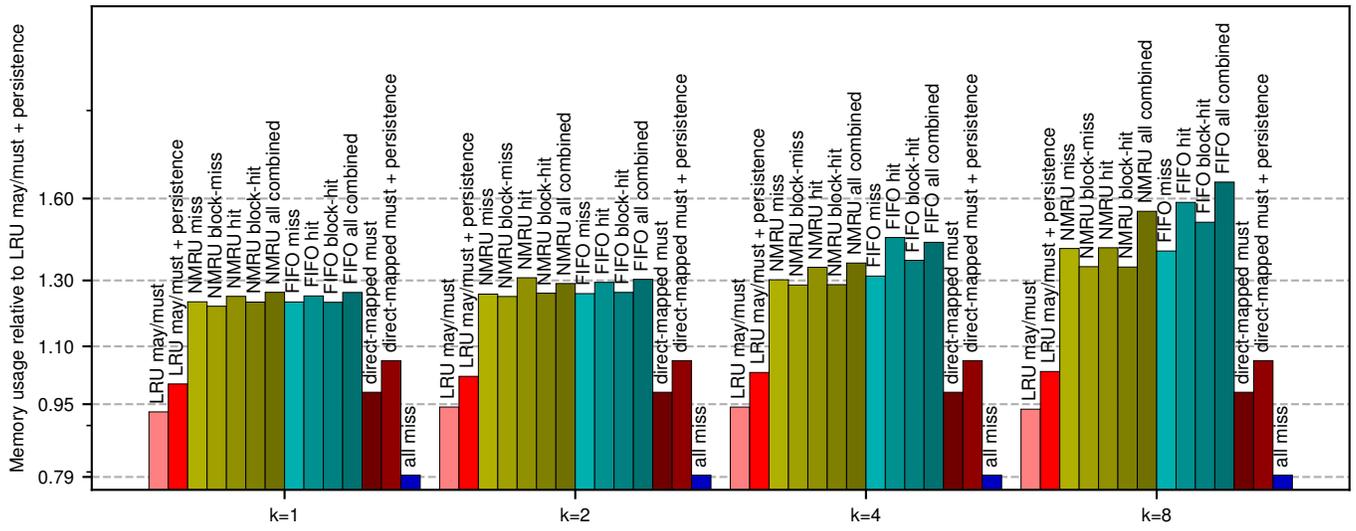
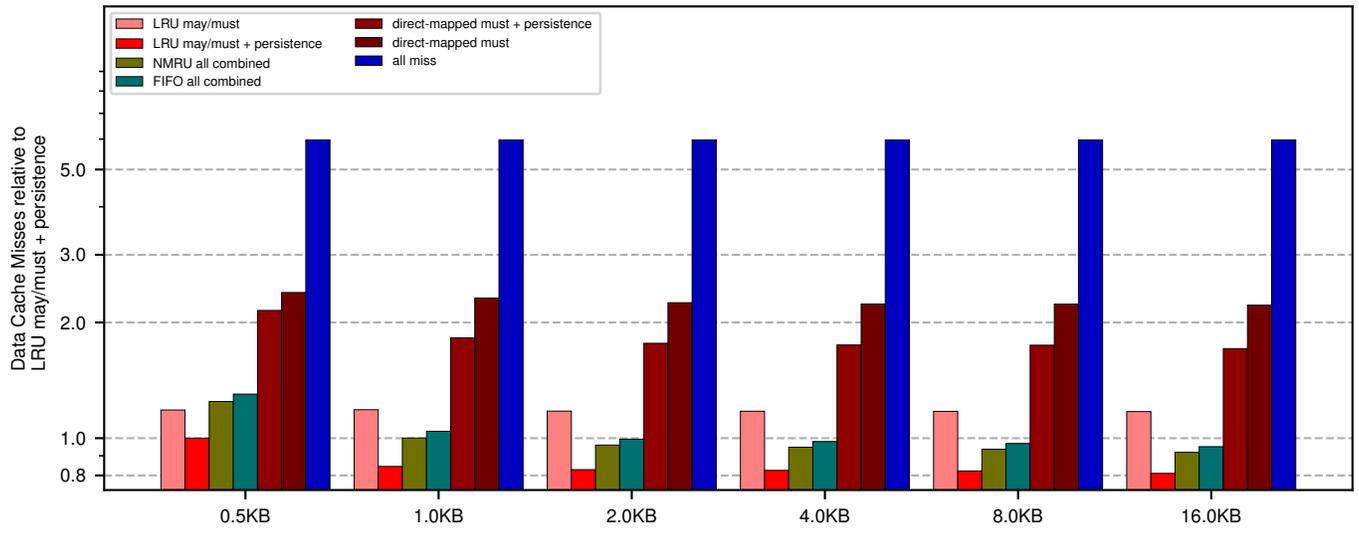
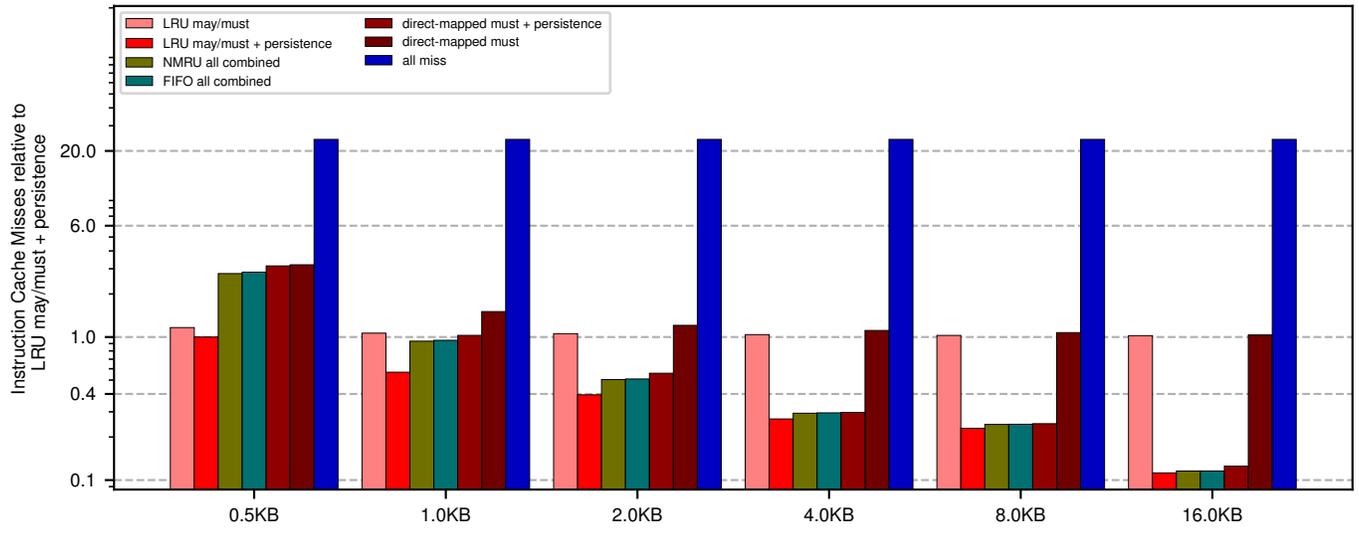
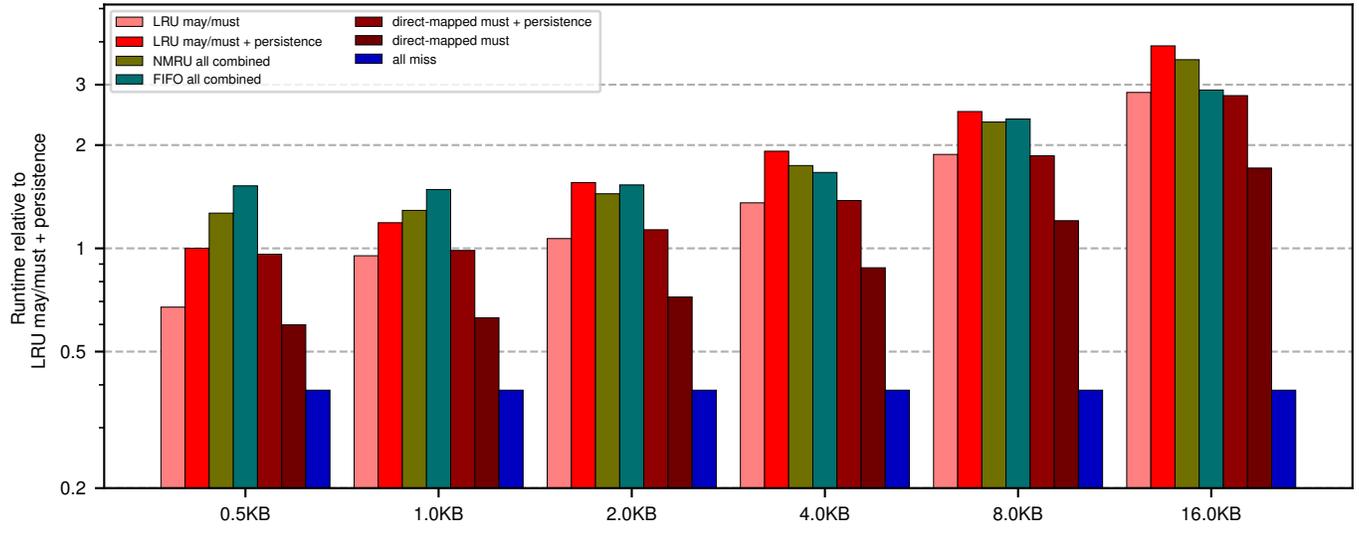
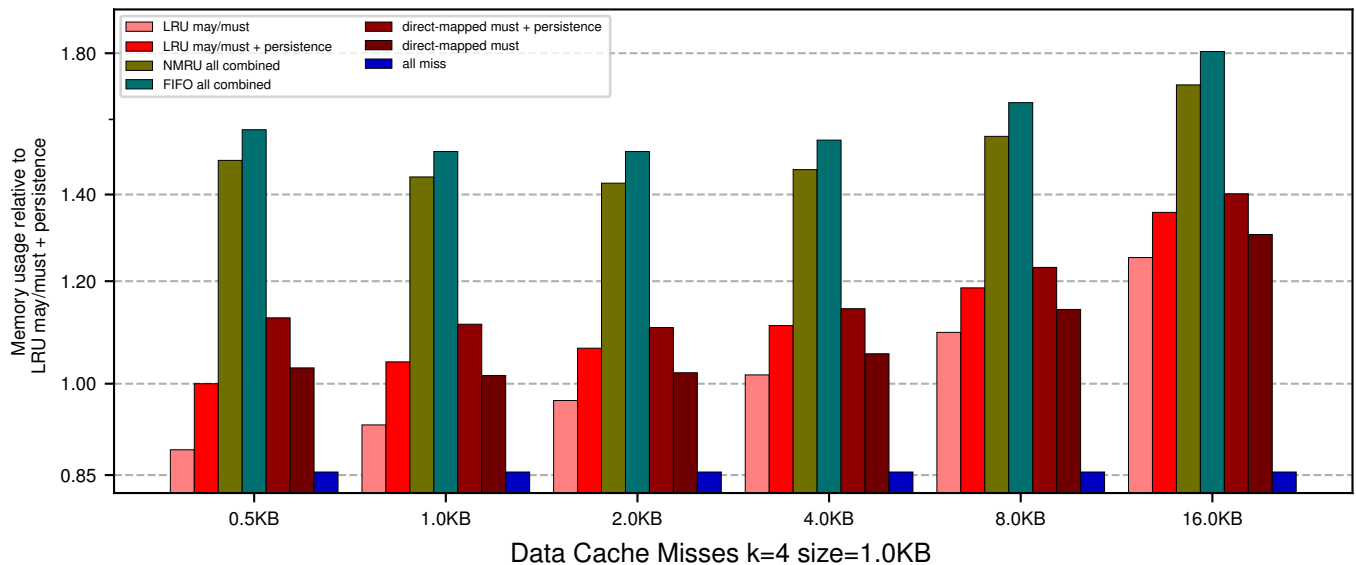
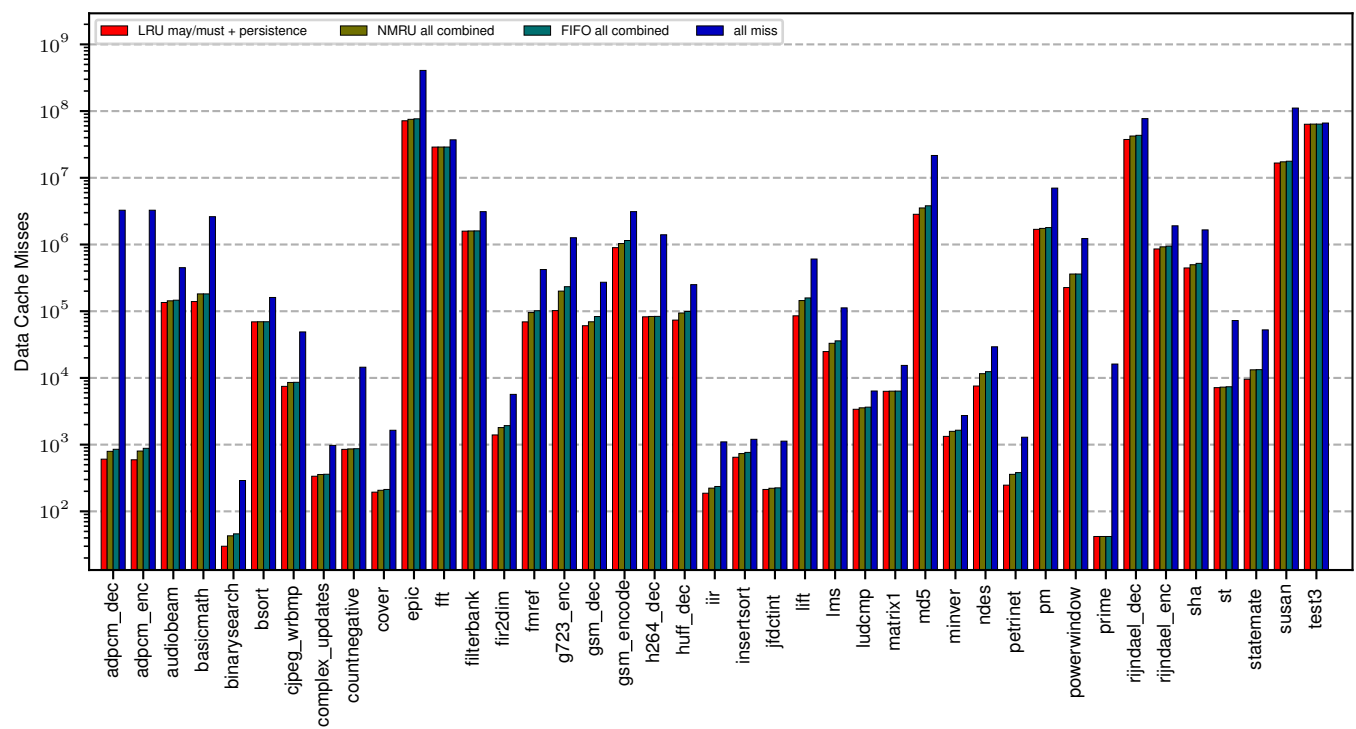
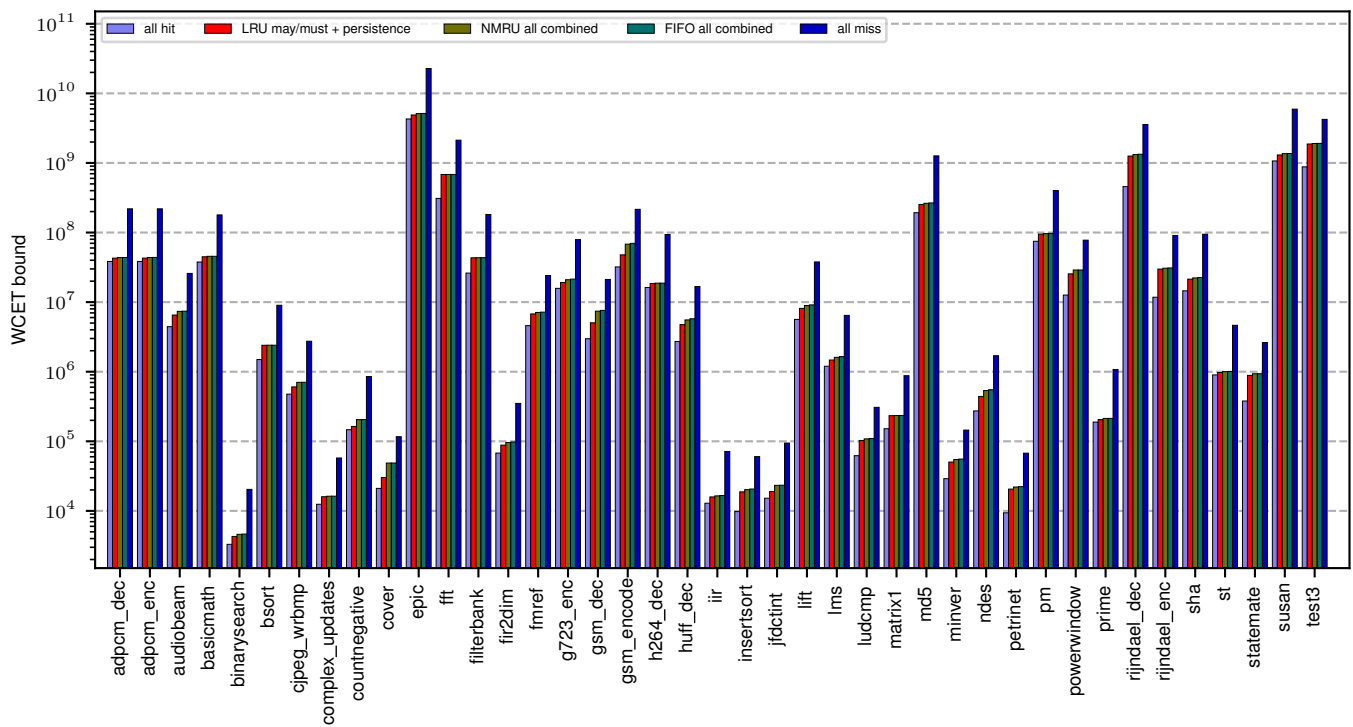
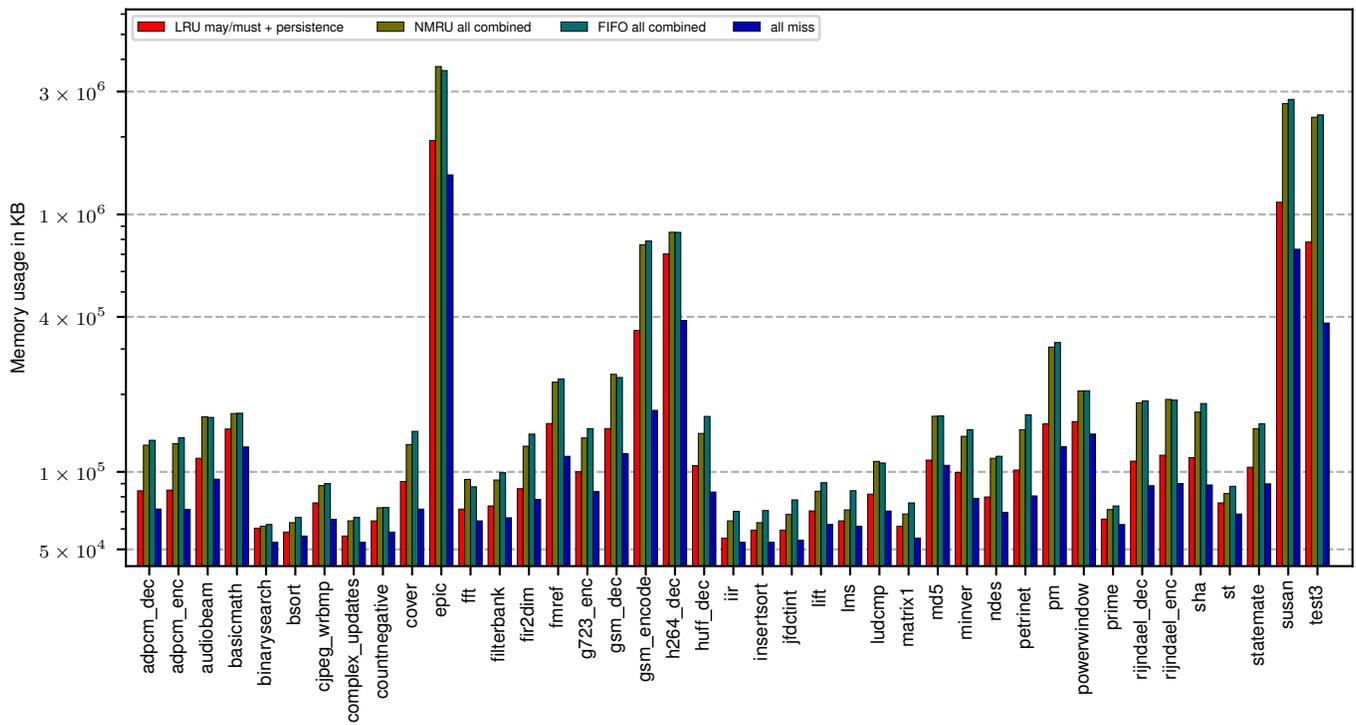
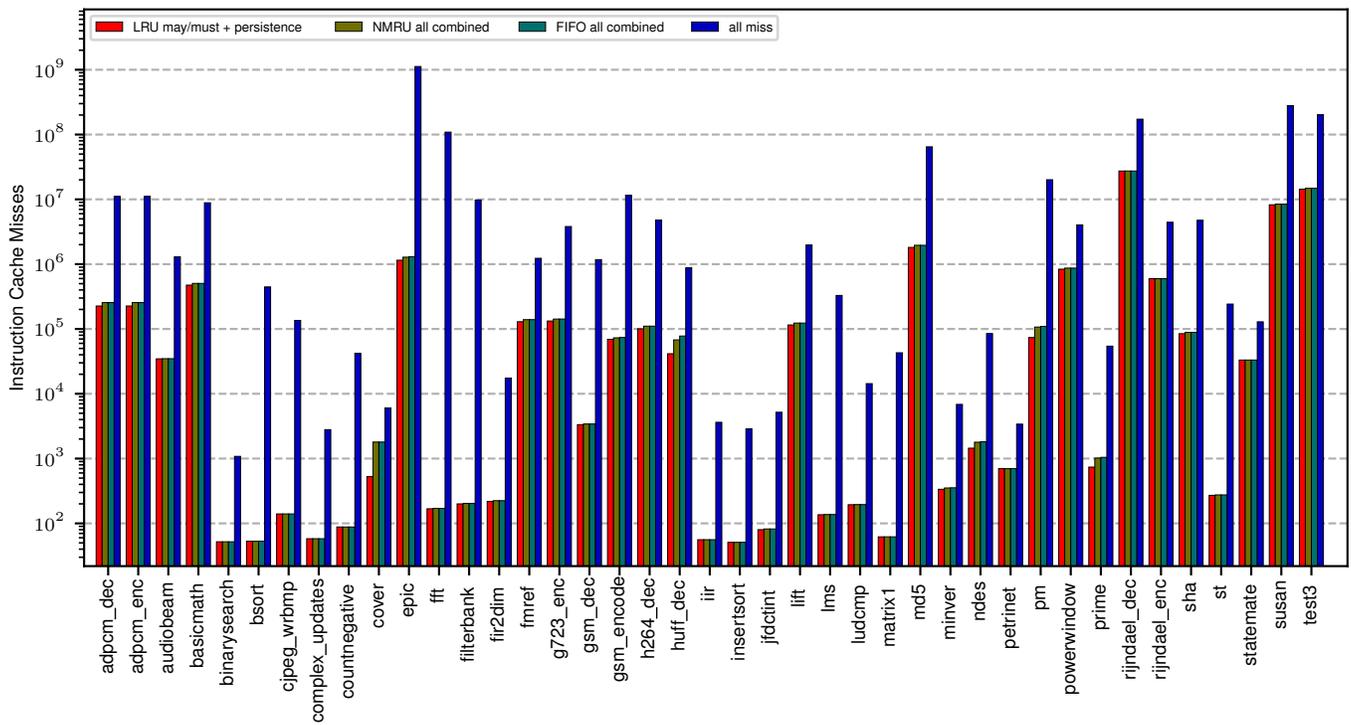
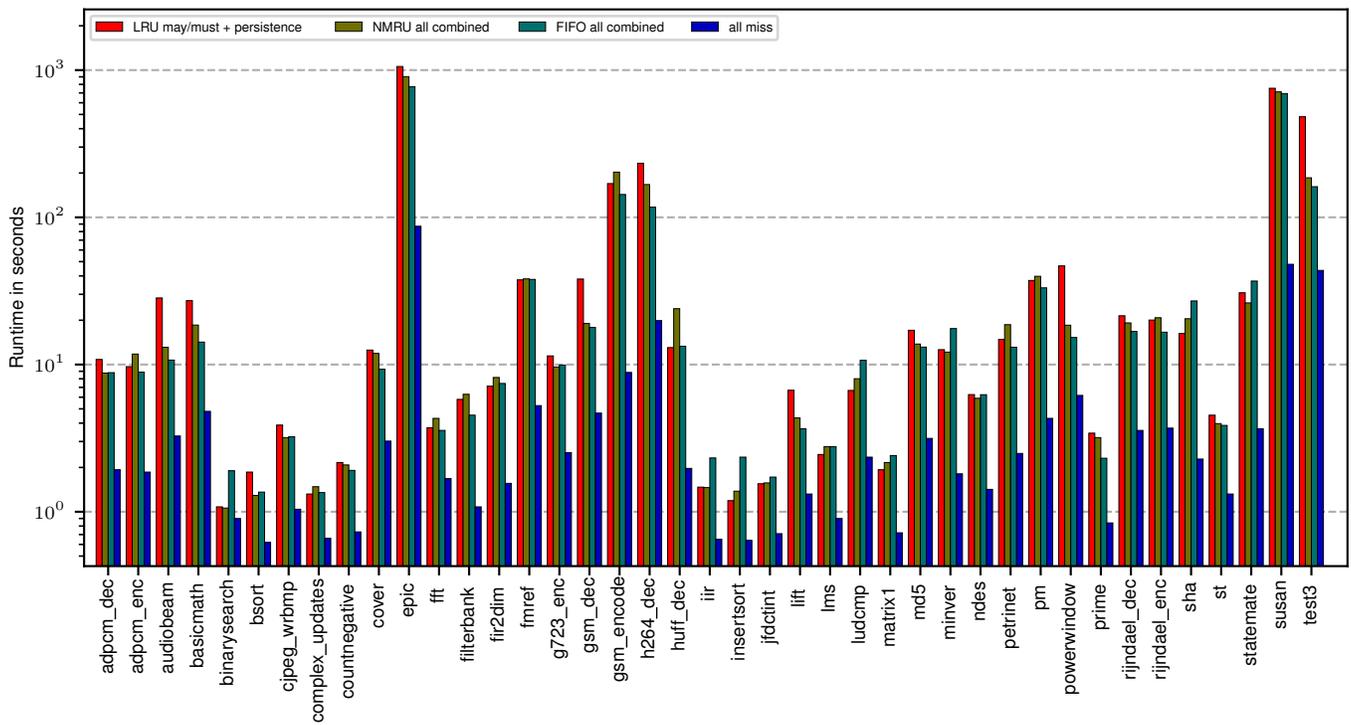
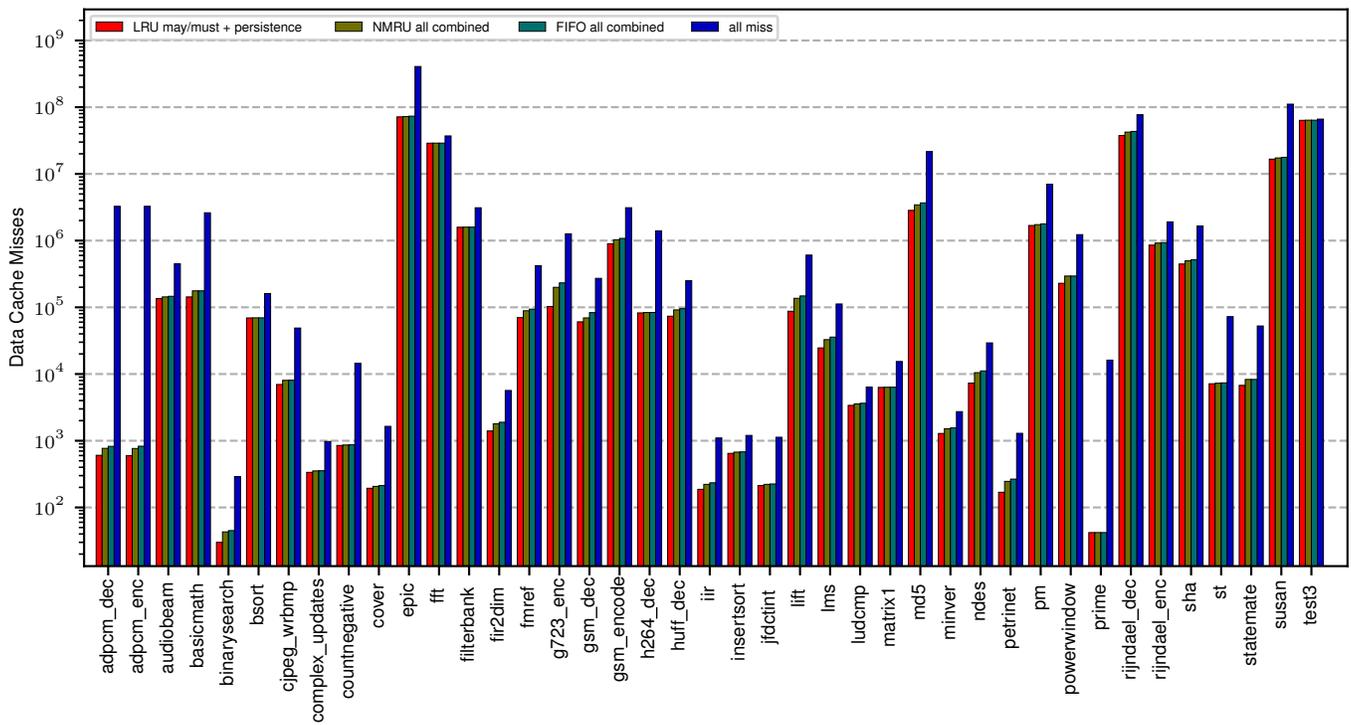
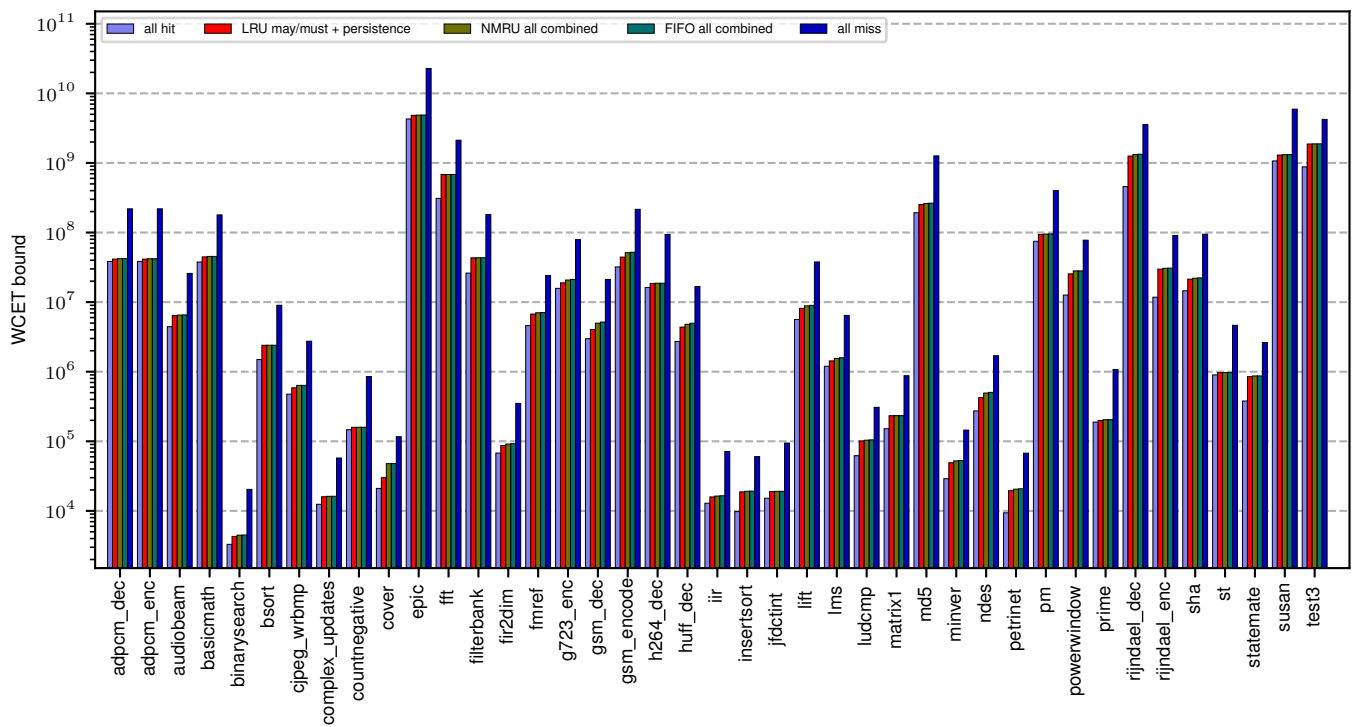
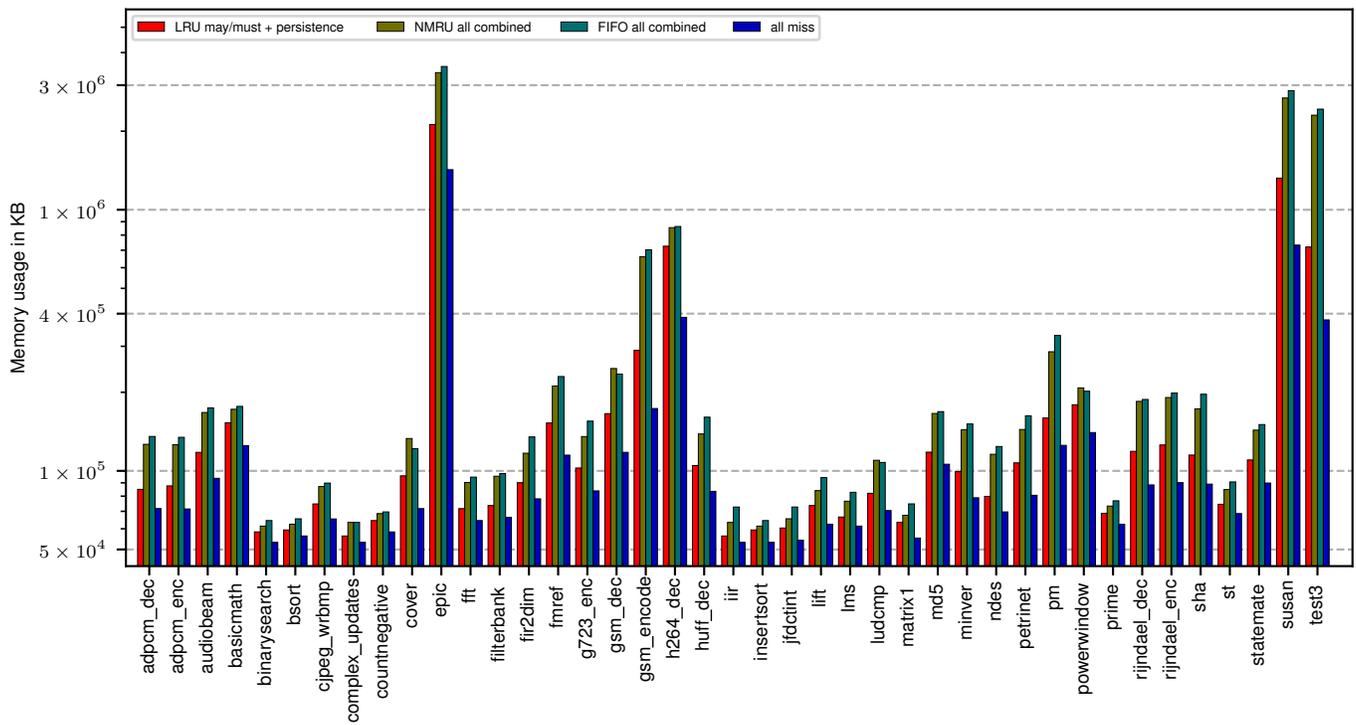
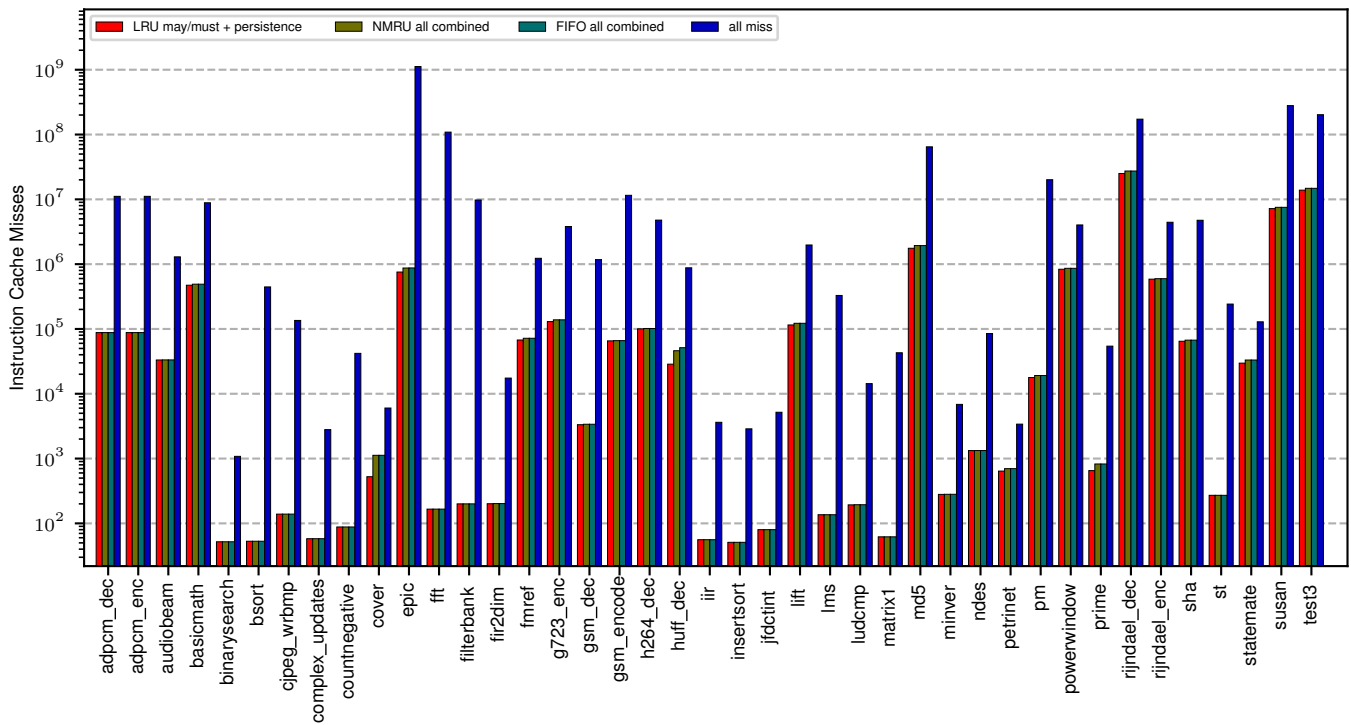
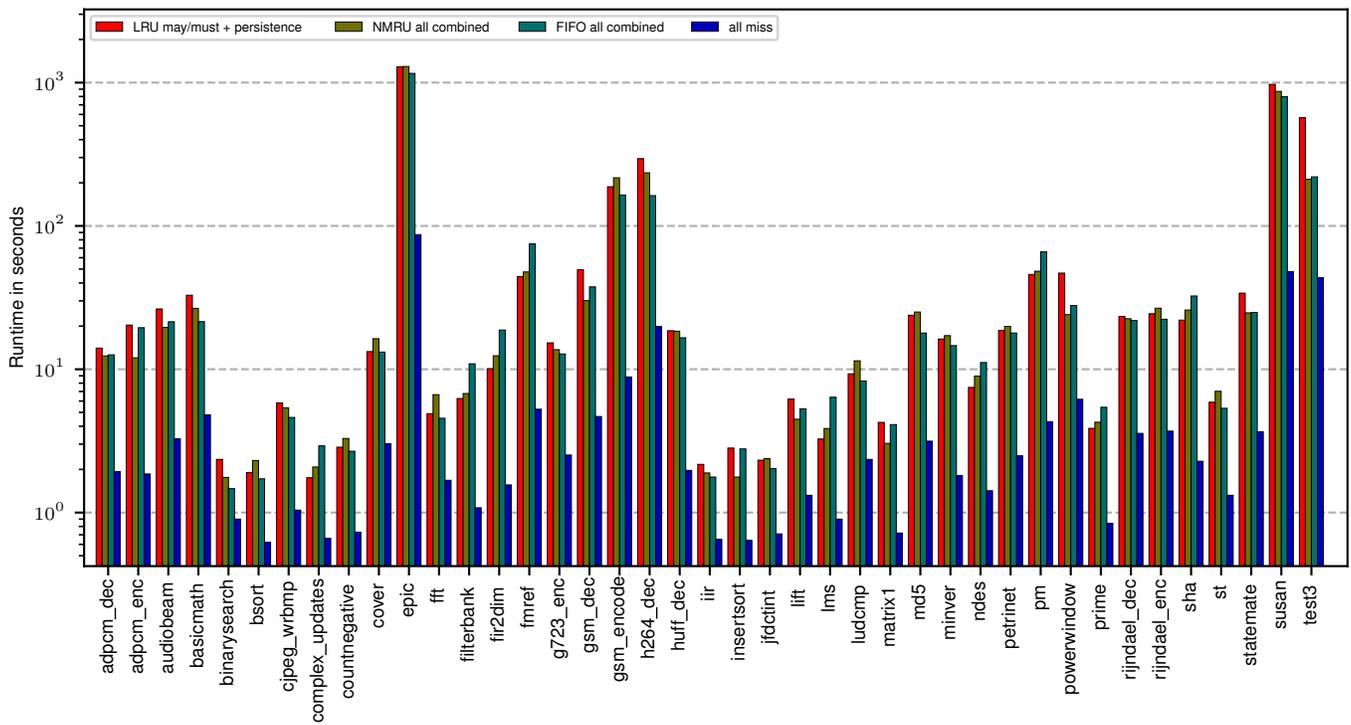
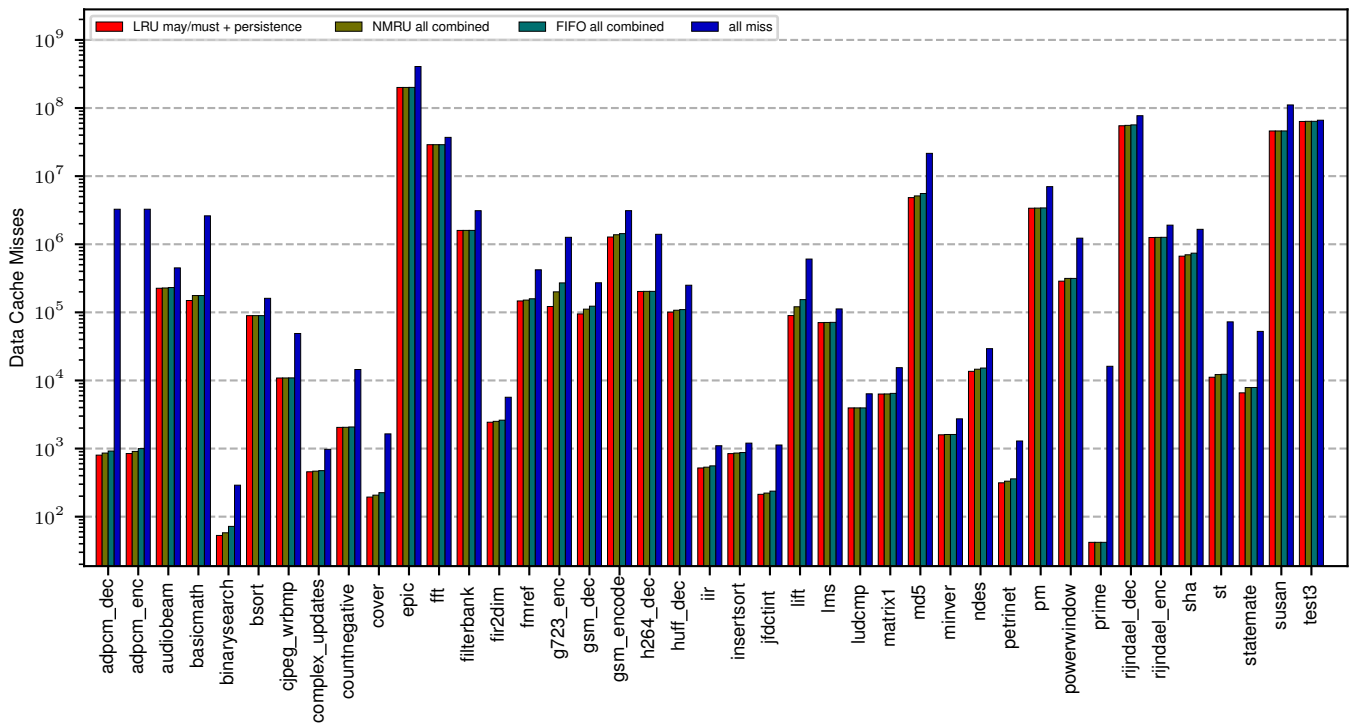
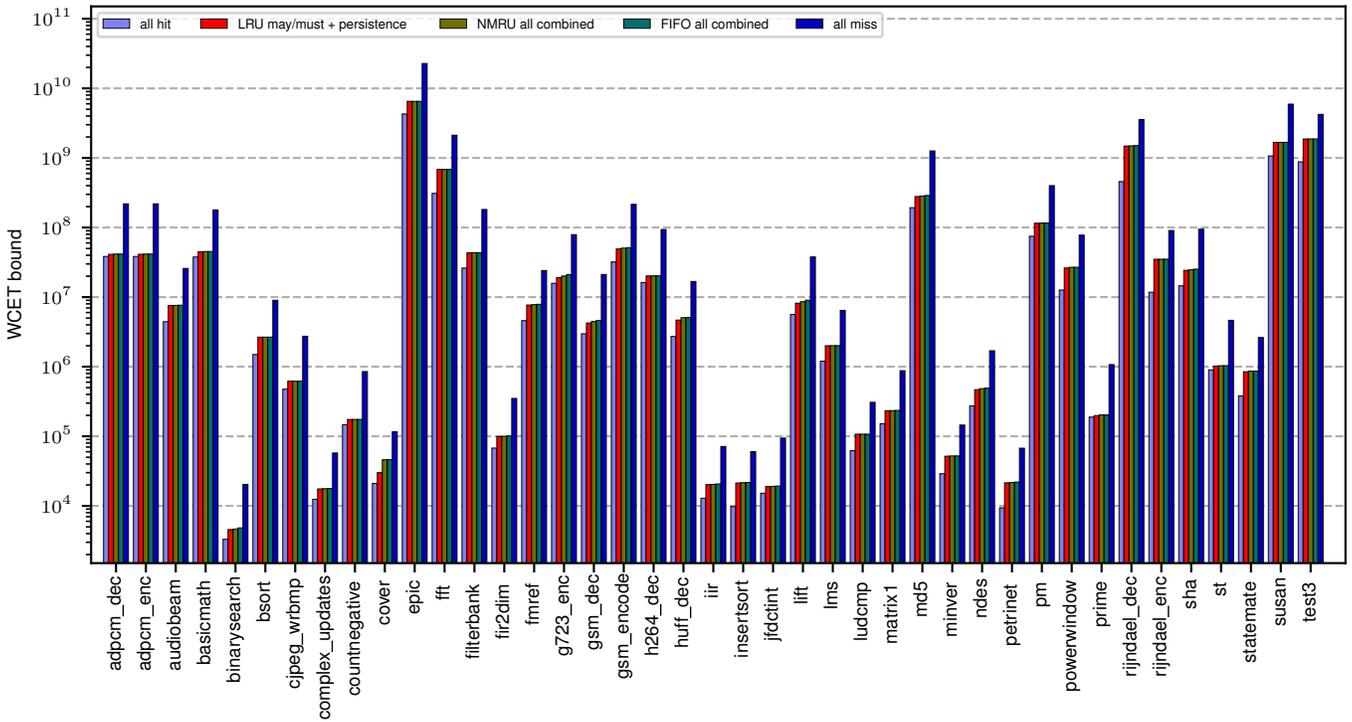
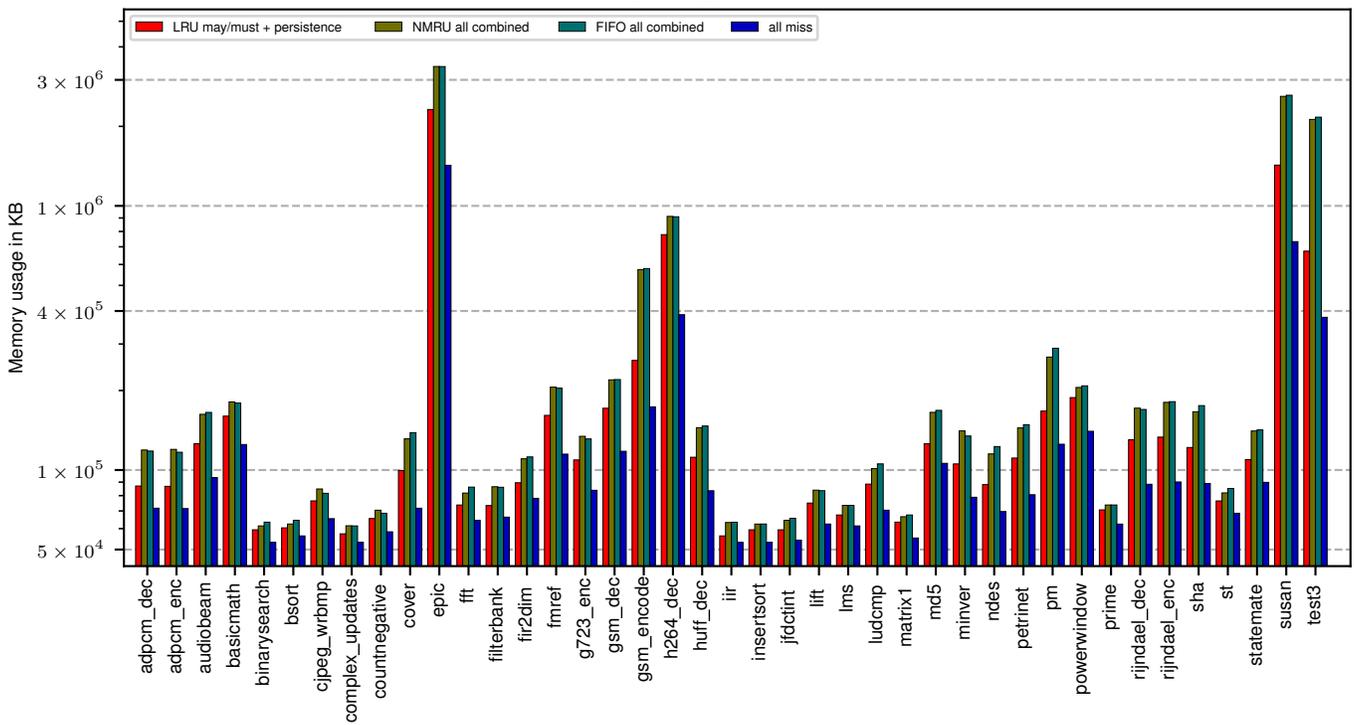
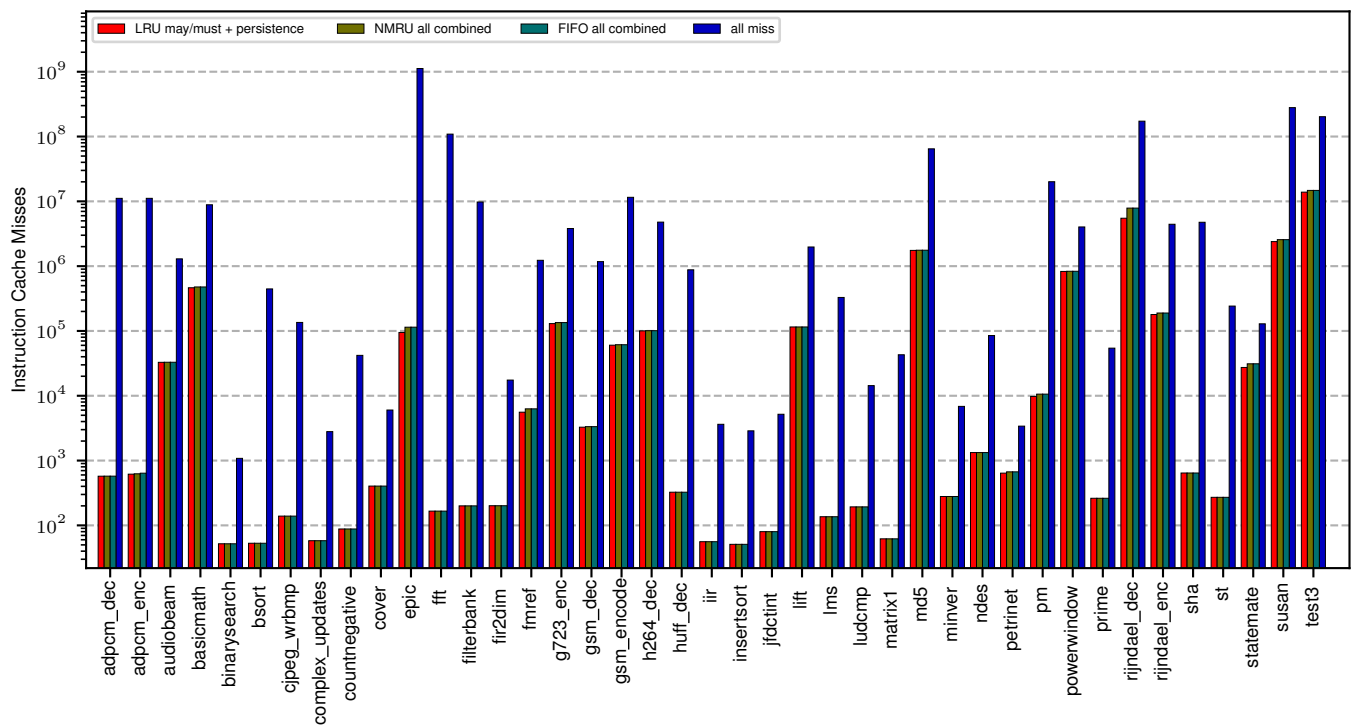
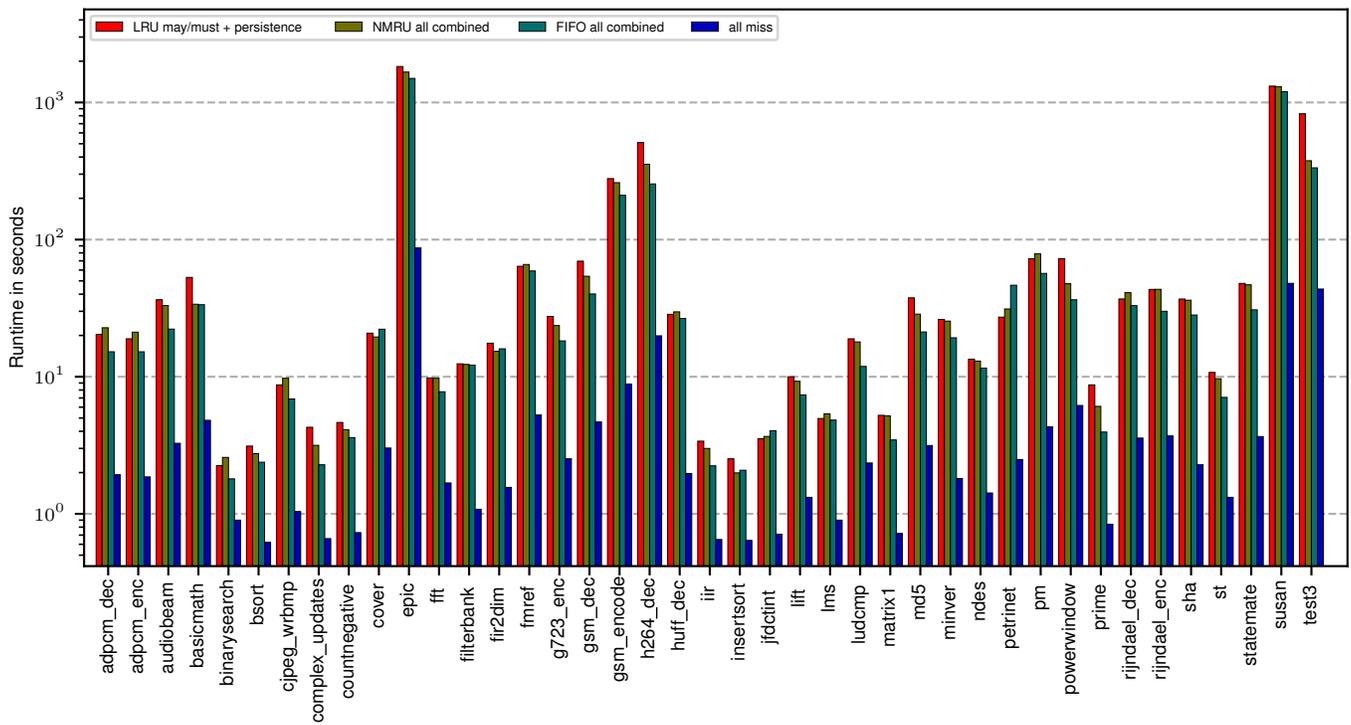
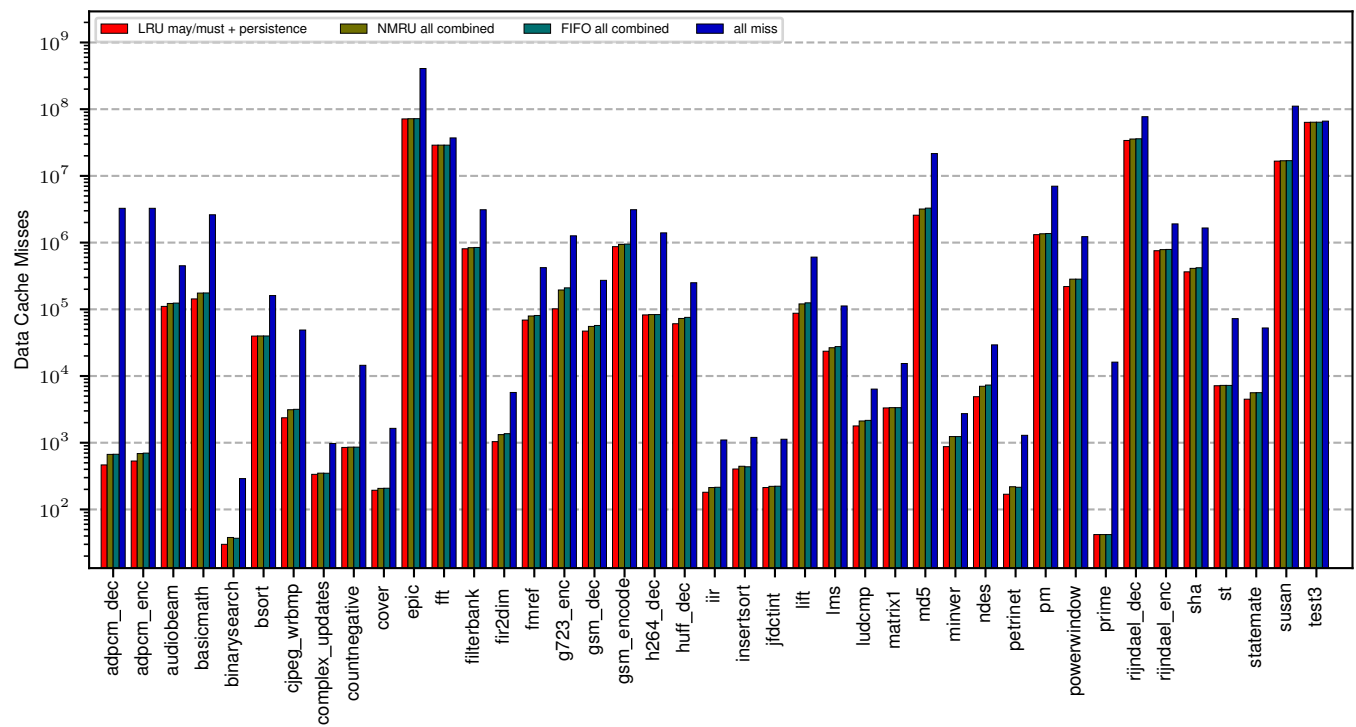
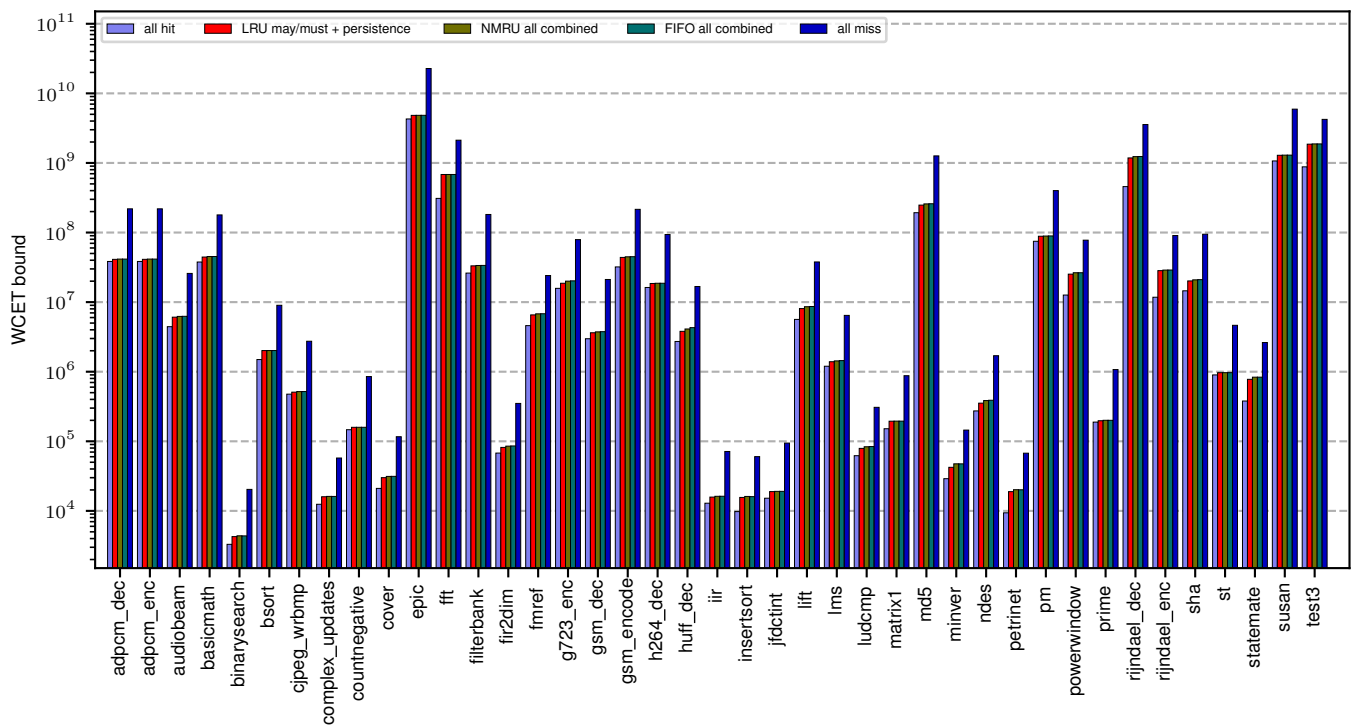
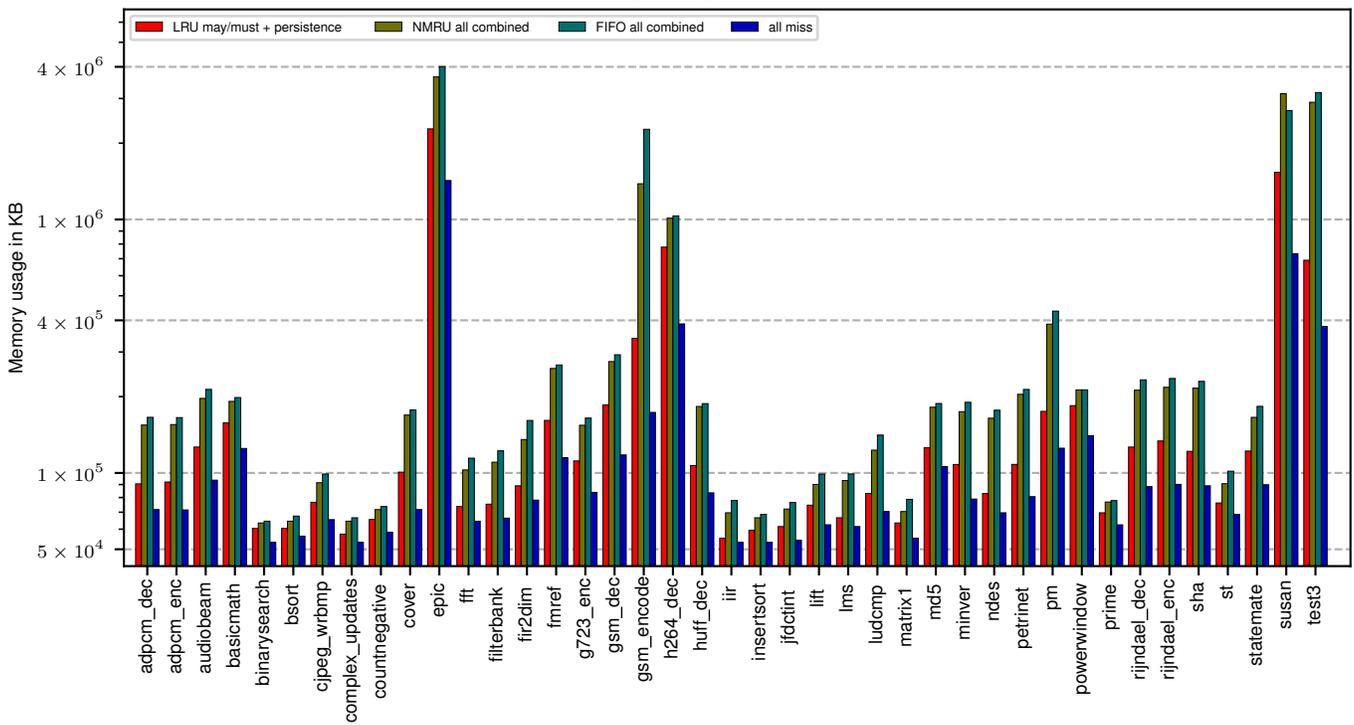
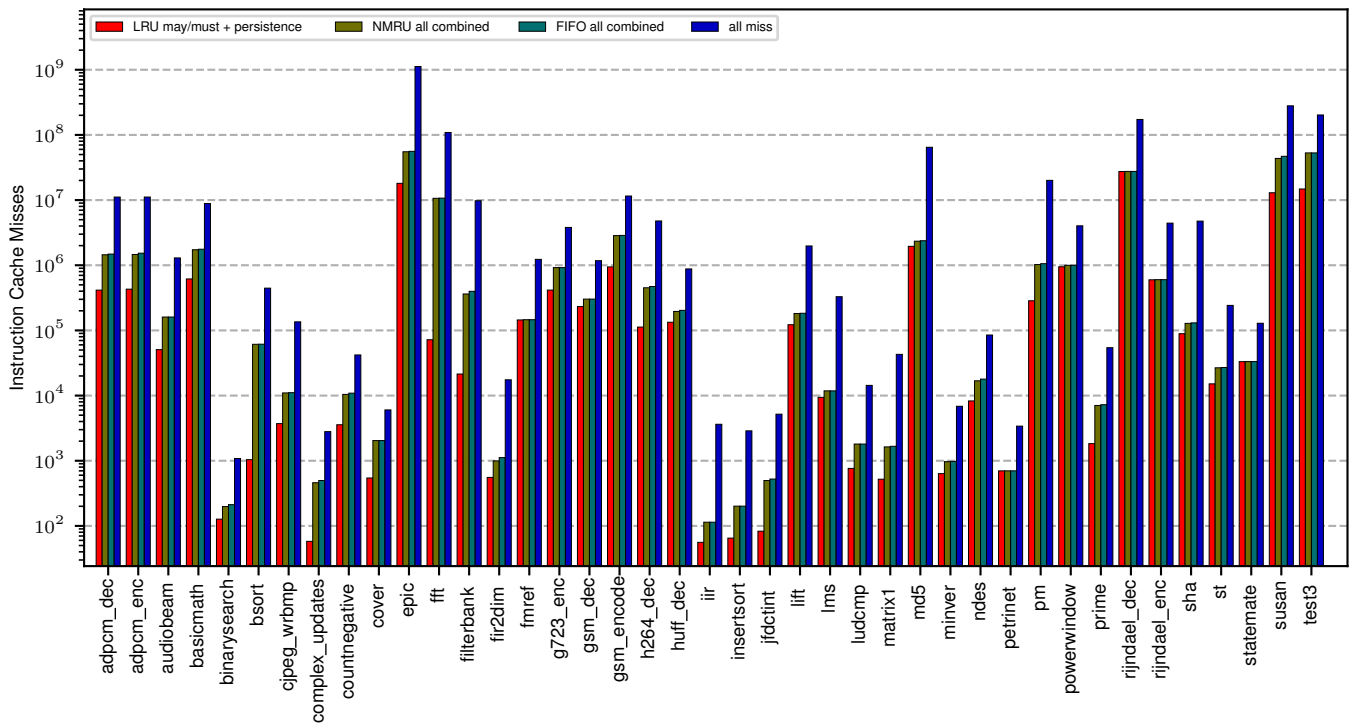
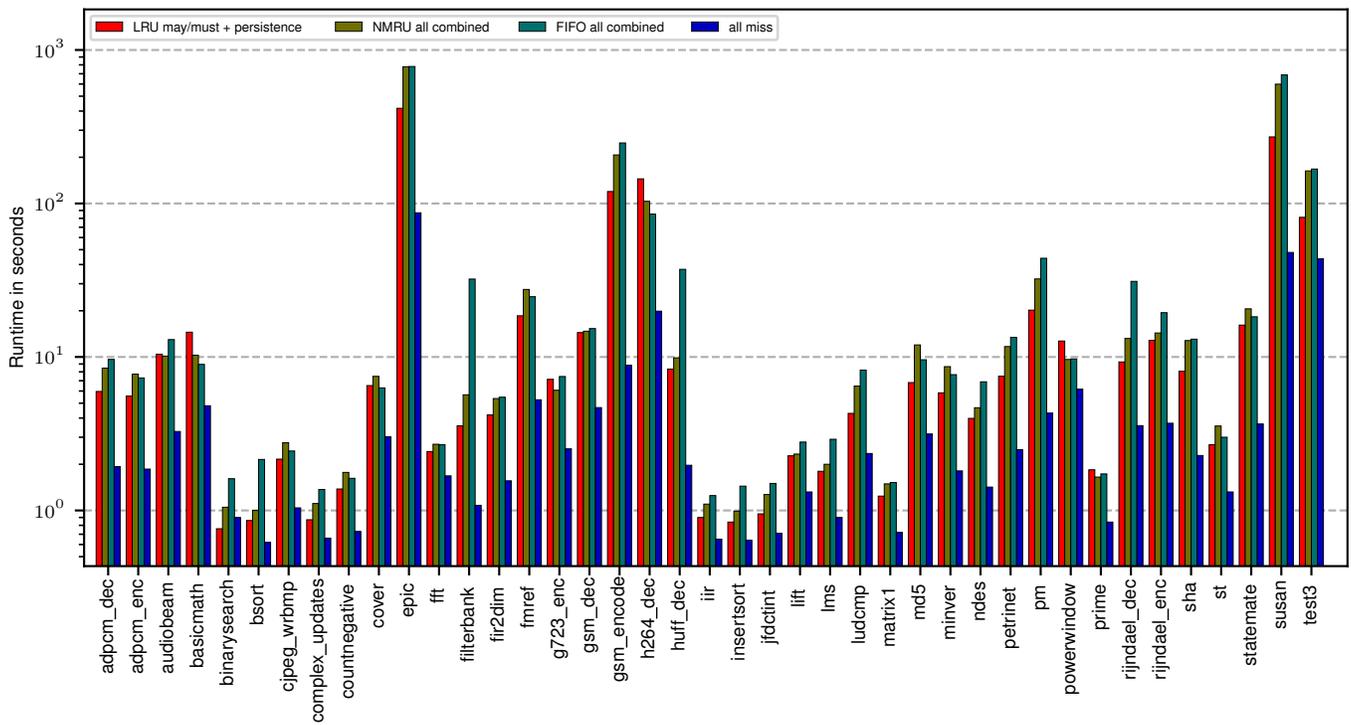
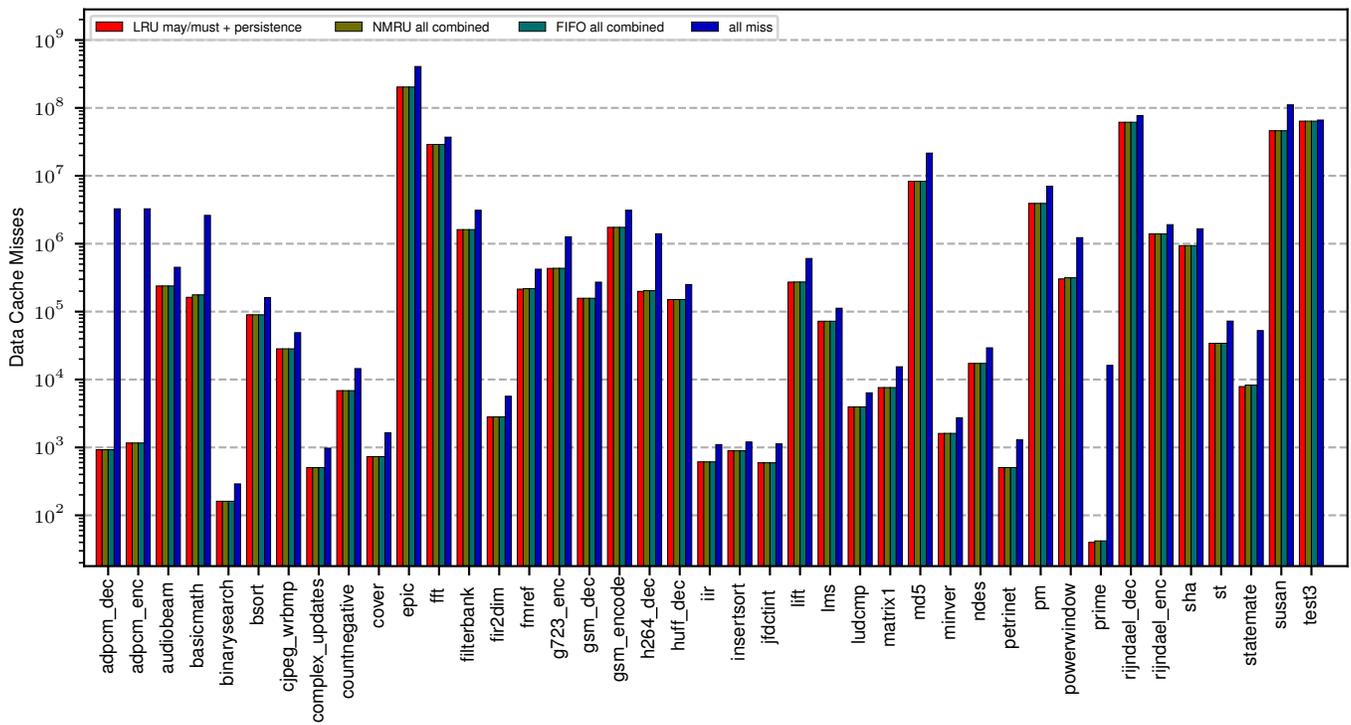
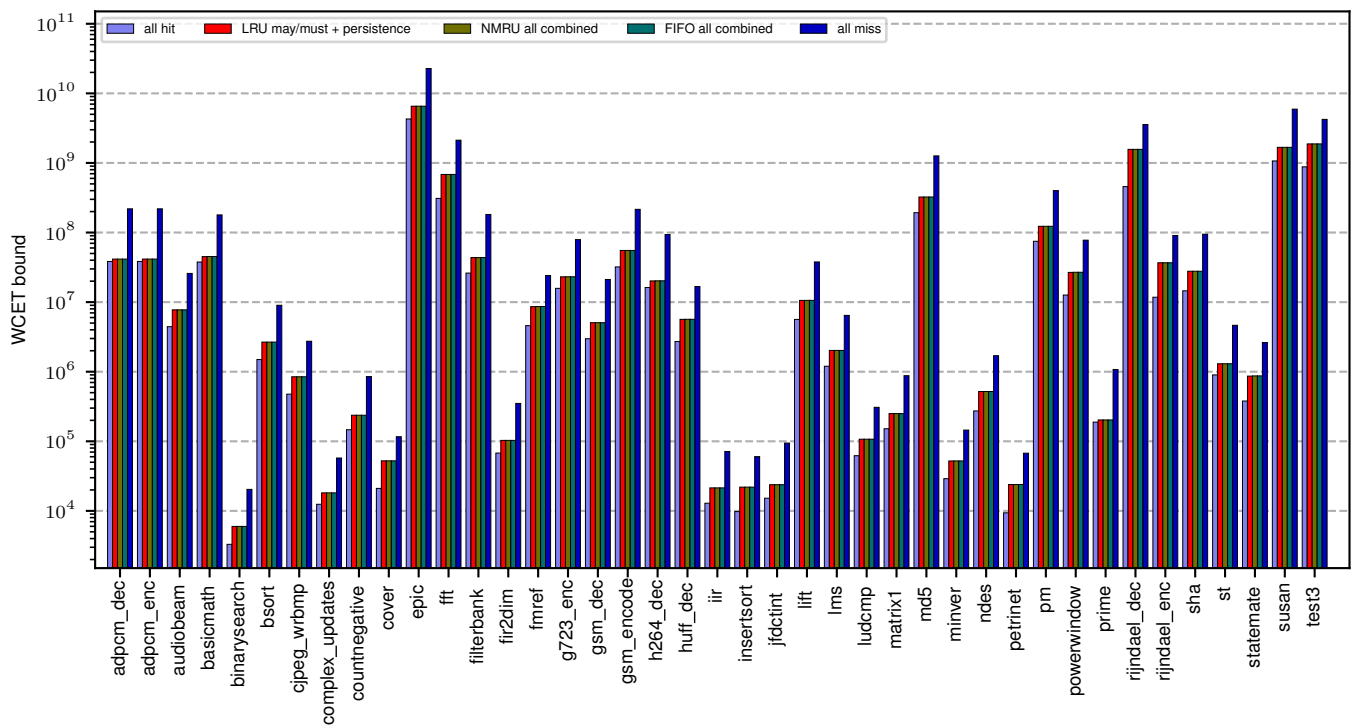
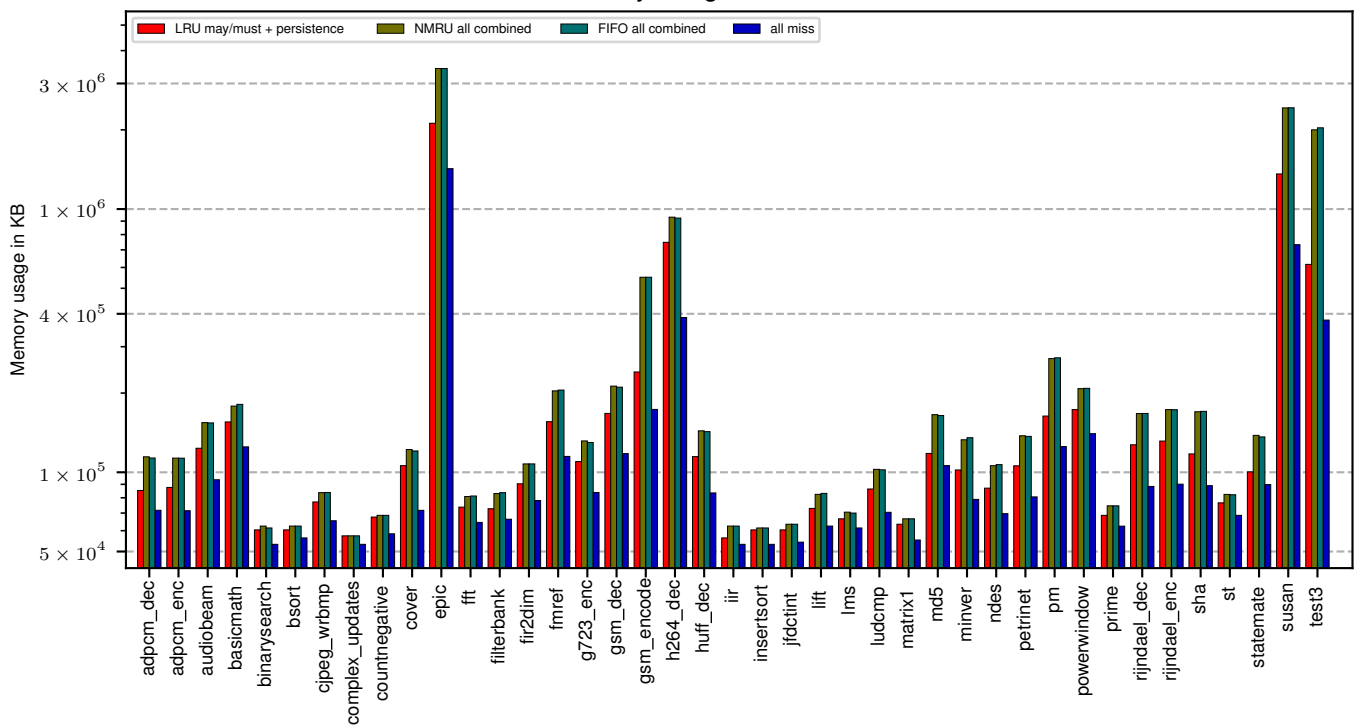
\end{document}